\documentclass[twoside]{article}

\usepackage[accepted]{aistats2025}
\usepackage[round]{natbib}

\usepackage[utf8]{inputenc}
\usepackage[T1]{fontenc}
\usepackage{lmodern}
\usepackage{textcomp}
\usepackage{microtype}
\usepackage{amsfonts}
\usepackage{nicefrac}
\usepackage{graphicx}
\usepackage{amsmath}
\usepackage{amsthm}
\usepackage{amssymb}
\usepackage{booktabs}
\usepackage{algorithm}
\usepackage{algpseudocode}
\usepackage{dsfont}
\usepackage{bm}
\usepackage{multirow}
\usepackage[dvipsnames]{xcolor}
\usepackage{subcaption}
\usepackage{enumitem}
\usepackage{url}
\usepackage{listings}

\usepackage[pagebackref,breaklinks,colorlinks]{hyperref}
\hypersetup{colorlinks=true, allcolors=blue}

\def\m#1{\texttt{#1}}
\newcommand{\ie}{\emph{i.e.}}
\newcommand{\eg}{\emph{e.g.}}

\newcommand\yes[1]{\textcolor{ForestGreen}{#1}}

\newcommand\na[1]{\textcolor{Gray}{#1}}

\newtheorem{theorem}{Theorem}
\newtheorem{proposition}{Proposition}
\newtheorem{corollary}{Corollary}
\newtheorem{remark}{Remark}
\newtheorem{definition}{Definition}
\newcommand{\equivto}{\stackrel{\scalebox{0.5}{$\triangle$}}{=}}

\newtheorem*{proposition1}{Proposition 1}
\newtheorem*{proposition2}{Proposition 2}
\newtheorem*{proposition3}{Proposition 3}

\begin{document}

\runningauthor{A. Timans, C.-N. Straehle, K. Sakmann, C. A. Naesseth, E. Nalisnick}

\twocolumn[

\aistatstitle{Max-Rank: Efficient Multiple Testing for Conformal Prediction}

\aistatsauthor{
    Alexander Timans$^{*}$ \\ UvA-Bosch Delta Lab \\ University of Amsterdam
    \And
    Christoph-Nikolas Straehle$^{\text{\textdagger}}$ \\ Bosch Center for AI \\ Robert Bosch GmbH
    \And 
    Kaspar Sakmann \\ Bosch Center for AI \\ Robert Bosch GmbH
    \AND
    Christian A. Naesseth \\ UvA-Bosch Delta Lab \\ University of Amsterdam
    \And 
    Eric Nalisnick \\ Department of Computer Science \\ Johns Hopkins University
}

\aistatsaddress{
} 
]

\begin{abstract}
    Multiple hypothesis testing (MHT) frequently arises in scientific inquiries, and concurrent testing of multiple hypotheses inflates the risk of Type-I errors or false positives, rendering MHT corrections essential. This paper addresses MHT in the context of \emph{conformal prediction}, a flexible framework for predictive uncertainty quantification. Some conformal applications give rise to simultaneous testing, and positive dependencies among tests typically exist. We introduce \m{max-rank}, a novel correction that exploits these dependencies whilst efficiently controlling the family-wise error rate. Inspired by existing permutation-based corrections, \m{max-rank} leverages rank order information to improve performance and integrates readily with any conformal procedure. We establish its theoretical and empirical advantages over the common Bonferroni correction and its compatibility with conformal prediction, highlighting the potential to strengthen predictive uncertainty estimates.
\end{abstract}

%%%%%%%%% BODY TEXT
\section{INTRODUCTION}
\label{sec:intro}

Multiple hypothesis testing (MHT) occurs whenever we attempt to draw conclusions over multiple statistical hypotheses at once \citep{shaffer1995multiple}. Such scenarios regularly occur across different scientific domains ranging from medicine and genomics \citep{ludbrook1994medicine, qu2010genetic, storey2003gene} to economics and psychology \citep{blakesley2009psych, list2019economics}. A key issue is that as the number of tests grows, so does the risk of Type-I errors -- falsely rejecting the null hypothesis -- which can mislead scientific inquiry and waste resources on spurious or incorrect findings \citep{benjamini2010simultaneous}. To mitigate this, a vast body of statistical research has focused on multiple testing corrections that permit valid inference even in such cases \citep{bender2001adjusting}.

In this work, we examine a specific instance of the multiple testing problem: predictive uncertainty quantification through \emph{conformal prediction} \citep{g.shafer2008, v.vovk2005}. Conformal prediction offers a statistically rigorous framework for uncertainty estimation in complex predictive models, such as deep neural networks \citep{angelopoulos2023gentle}. However, MHT challenges arise when multiple conformal procedures are run simultaneously (\eg, for multi-target regression \citep{messoudi2021copula}), necessitating a correction. Moreover, these procedures often exhibit positive dependencies \citep{s.bates2022}, rendering standard approaches like Bonferroni inefficient \emph{a priori}.

% author footnotes
\renewcommand{\thefootnote}{\fnsymbol{footnote}}
\footnotetext{$^{*}$Corresponding author: $<$\texttt{a.r.timans@uva.nl}$>$} 
\footnotetext{$^{\text{\textdagger}}$Significant contributions.}
\setcounter{footnote}{0}
\renewcommand{\thefootnote}{\arabic{footnote}}

Leveraging the natural connections between conformal prediction and permutation testing, we introduce a simple MHT correction that exploits positive dependencies among hypotheses when evident through rank ordering. Our method ensures strong control of the family-wise error rate (FWER), \ie, the probability of at least one false discovery. Closely related to the permutation-based corrections of \cite{westfall1993resampling}, it readily integrates with the conditions assumed by conformal prediction without imposing any additional assumptions. In summary, we contribute:
\begin{itemize}[leftmargin=13pt]
    \item \m{max-rank}, an efficient correction for MHT settings with positive dependency (\autoref{algo:max-rank}), and establish its connection to the multiple testing procedures of \cite{westfall1993resampling} (\autoref{methods:relation-to-westfall-young}).
    \item Prove that \m{max-rank} offers superior FWER control over the Bonferroni correction using multivariate copula arguments (\autoref{sec:theory}, \autoref{sec:experiments}). 
    \item Demonstrate both theoretically (\autoref{subsec:conf-max-rank-0}) and empirically (\autoref{subsec:conf-exp-reg}) \m{max-rank}'s suitability for conformal prediction settings with multiple testing.
\end{itemize}

\section{BACKGROUND \& RELATED WORK}
\label{sec:background}

We first discuss our multiple testing setting and Type-I error control, with a closer look at the Bonferroni correction. Then, we introduce the conformal prediction setting and its connection to hypothesis testing.

\textbf{Multiple testing in parallel and in series.} A useful distinction can be made between statistical testing \emph{in parallel} and \emph{in series} \citep{good1958significance}. When testing in parallel, multiple hypotheses are examined simultaneously. Frequently, conclusions are drawn on the basis of the same underlying evidence such as a single common dataset \citep{vovk2022admissible}, making independence assumptions harder to justify. This is our setting of interest, and the one underlying conformal prediction and any arising MHT issues (see \autoref{sec:conformal}). In contrast, testing in series suggests conclusions are drawn one after another, and the outcome of a test may determine whether and which subsequent hypotheses should be explored. For example, a clinical drug trial with multiple phases may only initiate a follow-up if results are satisfactory during the first phase \citep{bauer1991series}.

\textbf{Family-wise error rate control.} We can further distinguish among different types of Type-I error control while testing: control in the strong sense via the FWER, and in weaker forms via, \eg, the false discovery rate (see \autoref{subsec:app-fdr-control}). FWER control as the most stringent form aims to limit the probability of committing at least \emph{one} Type-I error across all considered hypotheses. If we test $m$ hypotheses and denote by $m_{\text{FP}}$ the number of falsely rejected null hypotheses (\ie, false positives), then FWER control at significance level $\alpha$ ensures that
\begin{equation}
    \mathbb{P}(m_{\text{FP}} \geq 1) \leq \alpha.
\label{eq:fwer}
\end{equation}
This strict error control is appropriate in settings where false discoveries may have strong implications, such as in drug testing or clinical trials. Since conformal prediction is oftentimes motivated from a safety perspective by its provided coverage guarantees and is applicable to such settings (\eg, \cite{lei2021conformal, candes2023conformalized, yin2024conformal}), FWER control is also the appropriate notion here.

\textbf{FWER control via the Bonferroni correction.} Common corrections with FWER control include Bonferroni \citep{sedgwick2012bonf}, Bonferroni-Holm \citep{aickin1996bonfholm}, or their \v{S}id\'{a}k-adjusted versions \citep{abdi2007sidak}, which all presume independence. Arguably the simplest and most widely employed, the classic Bonferroni correction adjusts each individual significance level by the (fixed) number of tested hypotheses, \ie, $\alpha_{\text{Bonf}} = \alpha/m$. Then, FWER control can be shown directly using an union bound argument (see \autoref{subsec:app-fwer-bonferroni}). However, the correction is well-known to be overly conservative under positive dependency of hypotheses \citep{vovk2022admissible, benjamini2010simultaneous}, and requires specifying the number of tests $m$ beforehand.

\textbf{Multiple testing under dependency.} Albeit MHT issues have been addressed as early as \cite{tukey1953problem}, the research on valid and efficient corrections under dependency remains remarkably active. \cite{westfall1993resampling} suggested an early resampling-based procedure for capturing data dependencies, while recent years have seen further work on arbitrary dependencies \citep{causeur2009control, sun2009large, leek2008general, liu2019cauchy, finner2017simes, gou2018hochberg, z.chi2022} and p-value merging \citep{wilson2019harmonic, vovk2022admissible, chen2023trade}. Indeed, a remark by \cite{z.chi2022} that \emph{``the gap between theory and practice is mildly unsettling''} suggests these questions remain far from fully solved.

\textbf{Multiple testing in conformal prediction.} Conformal applications necessitating some form of multiple testing correction are widespread, and include multi-step time series and trajectory estimation \citep{stankeviciute2021conformal, sun2023copula, xu2024conformal, lindemann2023safe, muthali2023multi}; multi-target classification and regression \citep{m.cauchois2021, fischefficient, messoudi2021copula, messoudi2022ellipsoidal, feldman2023calibrated}, among them conformal object detection (\cite{l.andeol2023a, mukama2024copula}, see \autoref{subsec:conf-exp-reg}); treatment analysis and screening \citep{lei2021conformal, yin2024conformal, jin2023selection, a.fisch2022}, and even as diverse as early-exiting and diffusion models \citep{t.schuster2021, teneggi2023trust}. 

Throughout, employed MHT corrections frequently include Bonferroni \citep{stankeviciute2021conformal, lindemann2023safe, muthali2023multi, l.andeol2023a, jin2023selection}, but can range from Copulas \citep{messoudi2021copula, sun2023copula} to Simes \citep{fischefficient} and other forms of sequential testing  \citep{lei2021conformal, yin2024conformal, m.cauchois2021, angelopoulos2021learn}, showcasing a variety of considered solutions. Naturally, the application-specific test design (in parallel \emph{vs.} in series) largely governs the suitability of employing particular corrections.

\begin{algorithm}[t]
\caption{Split conformal prediction (general case)}
\label{algo:split-cp}
\begin{algorithmic}[1]
    
    \State \textbf{Input:} Pre-trained predictor $\hat{f}$, calibration samples $\mathcal{D}_{cal}=\{(X_i,Y_i)\}_{i=1}^{n} \sim \mathcal{P}_{XY}$, test sample $(X_{n+1},Y_{n+1}) \sim \mathcal{P}_{XY}$, miscoverage rate $\alpha \in (0,1)$.
    
    \State \textbf{Output:} Prediction set $\hat{C}(X_{n+1})$ for test sample.
    
    \State \textbf{Procedure:} 

    \State Define a scoring function $s:\mathcal{X}\times\mathcal{Y}\rightarrow\mathbb{R}$ applied to $\mathcal{D}_{cal}$, resulting in a set of nonconformity scores $S = \{s(\hat{f}(X_i),Y_i)\}_{i=1}^{n} = \{s_i\}_{i=1}^{n}$. The score $s_i$ encodes a notion of dissimilarity or \emph{nonconformity} between the prediction $\hat{f}(X_i)$ and true value $Y_i$.
    
    \State Compute a threshold $\hat{Q}(1-\alpha; S) \equivto \hat{Q}_{1-\alpha}$ as the $\lceil (n+1)(1-\alpha)\rceil/n$-th empirical quantile of $S$.
    
    \State For a new test sample $(X_{n+1},Y_{n+1})$, a \emph{valid} prediction set satisfying \autoref{eq:conf-guarant} is given by $\hat{C}(X_{n+1})=\{y\in\mathcal{Y}: \, s(\hat{f}(X_{n+1}), \,y) \le \hat{Q}(1-\alpha; S)\}$.
    
    \State \textbf{End Procedure}
\end{algorithmic}
\end{algorithm}

\textbf{Conformal prediction.} Conformal prediction (CP) offers a distribution-free, \emph{post-hoc}, and finite-sample approach to predictive uncertainty quantification \citep{v.vovk2005, g.shafer2008}. The obtained uncertainty around a predictor's output is expressed as a prediction set, and provides an associated guarantee on coverage (\autoref{eq:conf-guarant}). The general algorithmic recipe for split CP \citep{h.papadopoulos2007}, its most popular variant, is outlined in \autoref{algo:split-cp}. Split CP requires a separate calibration dataset $\mathcal{D}_{cal}=\{(X_i,Y_i)\}_{i=1}^{n} \sim \mathcal{P}_{XY}$ for which dissimilarity values (so-called \emph{nonconformity scores}) between predicted and true outputs are computed. Given a new test sample $(X_{n+1},Y_{n+1}) \sim \mathcal{P}_{XY}$, a prediction set is then constructed by including any predictions that align with the previously observed scores on $\mathcal{D}_{cal}$, as governed by a conformal quantile $\hat{Q}_{1-\alpha}$ ensuring a coverage rate of $(1-\alpha)$ for true values. The approach is distribution-free since it does not impose any distributional assumptions. Instead, a key requirement is the \emph{exchangeability} of $\mathcal{D}_{cal} \,\cup\, \{(X_{n+1}, Y_{n+1})\}$, a relaxed \emph{i.i.d.} assumption suggesting permutation invariance of samples. It is \emph{post-hoc} because it can be used with any `black-box' pre-trained predictor without the need for training or internal access. And it is finite-sample because the obtainable coverage guarantee does not rely on asymptotics. Specifically, if CP's conditions are met, a probabilistic statement on inclusion of $Y_{n+1}$ in the prediction set $\hat{C}(X_{n+1})$ for a tolerated miscoverage rate $\alpha \in (0,1)$ is given as
\begin{equation} 
    \mathbb{P}(Y_{n+1} \in \hat{C}(X_{n+1})) \ge 1-\alpha.
\label{eq:conf-guarant}
\end{equation}
Importantly, the provided coverage guarantee is \emph{(i)} \emph{marginally} valid, \ie, it only holds on average across test and calibration data and may thus fail to cover for any particular observation; and \emph{(ii)} fails to appropriately cover if exchangeability is violated, such as under strong distribution shifts \citep{barber2023conformal}.

\textbf{Conformal prediction as hypothesis testing.} The CP procedure in \autoref{algo:split-cp} can also be interpreted from a hypothesis testing perspective \citep{shi2013applications, s.bates2022, v.vovk2005}. Given a candidate value $y \in \mathcal{Y}$, a statistical test for its inclusion in the prediction set can be formalized by the hypothesis pair $H_0: Y_{n+1} = y,\,\,H_1: Y_{n+1} \neq y$, with test statistic $s_{n+1} = s(\hat{f}(X_{n+1}), y)$ equating its nonconformity score. The set of calibration scores $S = \{s_1, \dots, s_n\}$ provides an empirical null distribution to compute a rank-based p-value $\hat{P}_{n+1}(y; S)$ as 
\begin{equation}
    \hat{P}_{n+1}(y; S) = \frac{|\{  i=1,\dots,n+1: s_i \ge s_{n+1} \}|}{n+1},
\label{eq:conformal-p-value}
\end{equation}
\ie, the fraction of samples which conform worse than $s_{n+1}$. A testing decision on not rejecting $H_0$ -- that is, inclusion in the prediction set -- is taken at significance level $\alpha$ as $\hat{P}_{n+1}(y; S) > \alpha$. The formulated testing approach translates to an instance of permutation testing: we are assessing if the observed nonconformity for candidate value $y$ does not stray overly far from previous observations, and the sample $(X_{n+1}, y)$ thus permutes with $D_{cal}$. Subsequently, we interpret running multiple conformal procedures at once as permutation testing \emph{in parallel}, and at identical level $\alpha$ across tests. 

\textbf{Equivalence of events.} By the prediction set construction outlined in \autoref{algo:split-cp} and the above testing perspective, a useful equivalence of events for a prediction set candidate value $y$ can be established as
\begin{align}
    y \in \hat{C}(X_{n+1}) &\Leftrightarrow s_{n+1} \le \hat{Q}(1-\alpha; S) \Leftrightarrow \hat{P}_{n+1}(y; S) > \alpha, \text{} \nonumber \\
    y \notin \hat{C}(X_{n+1}) &\Leftrightarrow s_{n+1} > \hat{Q}(1-\alpha; S) \Leftrightarrow \hat{P}_{n+1}(y; S) \le \alpha.
\label{eq:equiv-of-events}
\end{align}
Notably, determining set inclusion via the conformal quantile $\hat{Q}(1-\alpha; S)$ (abbreviated to $\hat{Q}_{1-\alpha}$) which ensures a coverage rate of $(1-\alpha)$ matches a permutation test using $\hat{P}_{n+1}(y; S)$ at significance level $\alpha$. In both cases, Type-I error control at the same rate $\alpha$ is achieved. We further comment on the quantile's properties in the following remarks:
\begin{remark}
    The Type-I error control property of $\hat{Q}_{1-\alpha}$ at level $\alpha$ can be expressed in terms of coverage, \ie, it is the quantile which ensures that at least a fraction $(1-\alpha)$ of values are below and thus covered by it. 
\label{remark:control-as-coverage}
\end{remark}
\begin{remark}
    The smallest possible quantile $\hat{Q}_{1-\alpha}$ which maintains Type-I error control at level $\alpha$ is tight in terms of coverage, \ie, it provides the smallest and thus most efficient conformal prediction sets. We may loosely interpret this as the matching permutation test ensuring highest statistical power.
\label{remark:equiv-of-events}
\end{remark}

\section{MULTIPLE TESTING CORRECTION VIA MAX-RANK}
\label{sec:methods}

We next formalize and describe our proposed multiple testing correction \m{max-rank}, and thoroughly connect it to corrections by \cite{westfall1993resampling} in \autoref{methods:relation-to-westfall-young}.

\textbf{Notation.} Matrices are denoted as upper-case bold ($\mathbf{M}$), vectors as lower-case bold ($\mathbf{v}$), scalars without bolding ($s$). A multivariate random variable may also be upper-case bold given the context ($\mathbf{X}$). An index set $\{1,\dots,n\}$ is abbreviated $[n]$ for any $n \ge 1$. Similarly, a sequence $x_1, \dots, x_n$ is shortened to $x_{1:n}$. The rank of element $x_i$ in a set $\{x_{1:n}\}$ of $n$ distinct elements (potentially by introducing jitter) is given by 
\begin{equation}
    r_i \equivto \text{rank}(x_i; \{x_{1:n}\}) = |\{j \in [n]: x_j \le x_i\}|,
\label{eq:rank-definition}
\end{equation}
\ie, the size of the set of elements below $x_i$. Finally, the $l^{\infty}$-norm is given by $ \lVert \mathbf{x} \rVert_{\infty} = \max_{1 \le i \le n}|x_i|$ for a discrete vector $\mathbf{x}$ of size $n$. See also \autoref{tab:notation}.

\subsection{The \m{max-rank} algorithm}
\label{subsec:methods-algorithm}

Consider running $m$ hypothesis tests in parallel, and for any particular testing dimension $k \in [m]$ the statistics forming an empirical null distribution are given by $\textbf{s}_k \in \mathbb{R}^n$. In the conformal setting, $\mathbf{s}_k$ equates the set of nonconformity scores on $\mathcal{D}_{cal}$. We now desire finding adjusted significance levels $\hat{\alpha}_k$ for all dimensions such that the FWER is globally controlled at target level $\alpha$. 

We algorithmically describe our proposed \m{max-rank} correction in \autoref{algo:max-rank} (exemplary code is given in Listing \ref{lst:maxrank}). First, the empirical nulls of equal size are collected in a matrix of test statistics $\mathbf{S} = [s_{i,k}] \in \mathbb{R}^{n \times m}$, where $i \in [n]$ and $k \in [m]$. The $m$ columns span the tested hypotheses, and the $n$ rows represent values in each null distribution. Next, we obtain a similar matrix containing the column-wise ranks of $\mathbf{S}$ as $\mathbf{R} = [r_{i,k}] \in \mathbb{N}^{n \times m}$, where also $i \in [n]$ and $k \in [m]$. By \autoref{eq:equiv-of-events}, testing at significance level $\alpha$ in each column is equivalent to computing the conformal quantile $\hat{Q}(1-\alpha; \mathbf{s}_k)$ in the space of test statistics. This in turn equates determining the value $s_{t,k}$, where $t = \hat{Q}(1-\alpha; \mathbf{r}_k)$ denotes a quantile in \emph{rank space} ensuring the desired error control (\autoref{remark:control-as-coverage}). The \m{max-rank} correction adjusts these rank quantiles for FWER control by forming a composite empirical null $\mathbf{r}_{\max}$ via the $l^{\infty}$-norm across tests, and using the rank $r_{\max} = \hat{Q}(1-\alpha; \mathbf{r}_{\max})$ to inform corrected significance levels $\hat{\alpha}_k$ for each hypothesis $k \in [m]$.

\begin{algorithm}[t]
\caption{Multiple testing correction via \m{max-rank}}
\label{algo:max-rank}
\begin{algorithmic}[1]
    
    \State \textbf{Input:} Hypothesis tests $k=1,\dots,m$ and associated empirical nulls $\mathbf{s}_1,\dots,\mathbf{s}_m$ of equal size $n$.
    
    \State \textbf{Output:} Adjusted quantiles with FWER control at level $\alpha$ as $(\hat{Q}_{1-\hat{\alpha}_1}, \dots, \hat{Q}_{1-\hat{\alpha}_m}) \in \mathbb{R}^m$.
    
    \State \textbf{Procedure:} 

    \State Collect empirical nulls in a matrix of test statistics $\mathbf{S} \in \mathbb{R}^{n \times m}$, and their respective column-wise ranks in a matrix of rank statistics $\mathbf{R} \in \mathbb{N}^{n \times m}$.

    \State Apply the $l^{\infty}$-norm row-wise to the ranks. Let $\mathbf{r}_i = (r_{i,1} \, \cdots \, r_{i,m})$ be the $i$-th row of $\mathbf{R}$, then

    $\mathbf{r}_{\max} = (\lVert \mathbf{r}_1 \rVert_{\infty}, \dots, \lVert \mathbf{r}_n \rVert_{\infty})^T = (\underset{1 \le k \le m}{\max} |r_{1,k}|, \cdots, \underset{1 \le k \le m}{\max} |r_{n,k}|)^T \in \mathbb{N}^{n}.$

    \State Compute the rank $r_{\max} = \hat{Q}(1-\alpha; \mathbf{r}_{\max})$ as the $\lceil (n+1)(1-\alpha)\rceil/n$-th empirical quantile of $\mathbf{r}_{\max}$.

    \State Sort $\mathbf{S}$ column-wise ascending such that $s_{1,k} < \dots < s_{n,k}$ for every $k \in [m]$.

    \State Select the test statistic $s_{r_{\max},k}$ at rank $r_{\max}$ column-wise in $\mathbf{S}$ as the adjusted quantile $\hat{Q}_{1-\hat{\alpha}_k}$ for every test dimension $k \in [m]$.
    
    \State \textbf{End Procedure}
\end{algorithmic}
\end{algorithm}

\textbf{Intuition.} Our \m{max-rank} correction intuitively collapses multiple empirical nulls into a single `worst-case' null for which the error-controlling quantile is computed. Thus, any potential dependencies across test statistics which are reflected in the rank ordering are efficiently aggregated in the `maximum rank' statistic $\mathbf{r}_{\max}$, and subsequently the selected rank quantile $r_{\max}$. By virtue of applying the $l^{\infty}$-norm, $r_{\max}$ ensures that adjusted significance levels obey FWER control. 

If the empirical nulls are uncorrelated and the tested hypotheses can thus be considered independent, no benefit is obtained from \m{max-rank}'s aggregation step. We prove that in such cases our approach performs \emph{at least} on par with the Bonferroni correction (\autoref{sec:theory}), and never worse. On the other hand, if the tests are perfectly correlated each empirical null exhibits identical rank ordering, and \m{max-rank} will provide the optimal quantile solution for each individual testing dimension. For any cases with positively dependent tests inbetween the two extremes, our approach ensures superior performance over Bonferroni, and scales well even as the number of hypotheses $m$ grows (see \autoref{sec:experiments}). We highlight that \emph{no additional assumptions} beyond those already underlying CP are required for the use of \m{max-rank}.

\subsection{Relation to Westfall \& Young corrections}
\label{methods:relation-to-westfall-young}

A strong connection exists between \m{max-rank} and the \m{max-T} and \m{min-P} resampling-based MHT corrections of \cite{westfall1993resampling}, which offer similar FWER control. The \m{max-T} method provides adjusted p-values by comparing test statistics to the distribution of \emph{maximum} statistics across hypotheses (\autoref{algo:max-t}), while the \m{min-P} method adjusts p-values based on the \emph{minimum} observed p-values after resampling (\autoref{algo:min-p}). In both cases, existing dependency structure between tests is leveraged through repeated data resampling. 

Exploring these two corrections through the lens of our conformal prediction setting, we find that \m{max-rank} can be cast as a procedure that \emph{(i)} naturally extends \m{max-T} to the rank space, and \emph{(ii)} describes a particular instance of \m{min-P} arising from our problem setting. As such, we argue that \m{max-rank} exhibits the advantages of both methods by permitting its direct application to obtained test statistics as in \m{max-T}, whilst maintaining scale invariance as in \m{min-P}. We next relate to each correction in more detail.

\textbf{Relation to \m{max-T}.} Our conformal setting directly provides empirical null distributions for every hypothesis by use of the calibration set. The computed nonconformity scores on $\mathcal{D}_{cal}$ leverage available labels to produce statistics known to satisfy the tested null $H_0: Y_{n+1} = y$, and are thus interpretable as an analog to the required ``data resampling under the null'' \citep{westfall1993resampling}. Similarly, CP's exchangeability condition aligns with the exchangeability underlying the resampling step. It follows that the adjusted p-values in \autoref{algo:max-t} match the conformal p-values in \autoref{eq:conformal-p-value} simply by replacing the score set $S$ with a \emph{maximum} score set $S_{\max}$, \ie, replacing $\hat{P}_{n+1}(y; S)$ with $\hat{P}_{n+1}(y; S_{\max})$\footnote{And correcting for $(n+1)$ samples instead of $n$.}. Instead of adjusting the p-values as in \autoref{algo:max-t} we may also directly target an adjustment of the conformal quantile (\autoref{eq:equiv-of-events}), which translates to computing $\hat{Q}(1-\alpha; S_{\max})$ instead of $\hat{Q}(1-\alpha; S)$. Comparing this to \m{max-rank}, we notice that the difference lies in its direct application to the maximum statistics, rather than translating them into rank space first (\ie, using $\hat{Q}(1-\alpha; \mathbf{r}_{\max})$). This means that \m{max-T} is sensitive to scale and requires test statistics across hypotheses to be directly comparable. Interestingly, such a ``maximum of scores'' approach has been intuitively employed by researchers in conformal prediction, \eg, \cite{l.andeol2023a, t.schuster2021, a.fisch2022}, but never been recognized as an application case of \cite{westfall1993resampling}.

\textbf{Relation to \m{min-P}.} By the same resampling interpretation as above, unadjusted p-values for each hypothesis in \autoref{algo:min-p} can be computed via conformal p-values (\autoref{eq:conformal-p-value}). Using the rank definition in \autoref{eq:rank-definition}, a given p-value $\hat{P}_{n+1}(y; S)$ can be expressed in terms of score ranking as $((n+1) - \text{rank}(s_{n+1}; S))/(n+1)$\footnote{Observe that the p-value in \autoref{eq:conformal-p-value} counts the fraction of values \emph{above} $s_{n+1}$, whereas the rank counts values \emph{below} it.}. Thus, computing the minimum p-values in \m{min-P} can be interpreted as the maximum over inverted and normalized ranks. Via \autoref{eq:equiv-of-events} we again translate adjusted p-values to a quantile adjustment using $\hat{Q}(\alpha;\mathbf{p}_{\min})$. Note that this quantile is now taken over p-values and only permits testing in p-value space. In contrast, \m{max-rank} keeps the simplicity of operating directly on statistics, facilitating the interpretation of testing decisions.

\section{THEORETICAL RESULTS}
\label{sec:theory}

We next briefly describe our theoretical results for \m{max-rank}, deferring details and proofs to \autoref{app:math}. The multiple testing problem is first rephrased in terms of multivariate copulas, and our two key results state that \emph{(i)} \m{max-rank} provides a valid MHT correction with FWER control, and \emph{(ii)} is assured to perform better than Bonferroni under positive dependency. 

\textbf{Recasting the multiple testing problem (\autoref{subsec:methods-copula-formulation}).} Consider an $m$-dimensional random variable $\mathbf{X} \in \mathbb{R}^m$ and denote its c.d.f. as $F_{\mathbf{X}}(\mathbf{x})$. Using the interpretation in \autoref{remark:control-as-coverage}, determining adjusted significance levels $(\hat{\alpha}_1, \dots, \hat{\alpha}_m)$ with joint Type-I error control at level $\alpha$ is then equivalent to finding a vector $\mathbf{x}_* \in \mathbb{R}^m$ satisfying $F_{\mathbf{X}}(\mathbf{x}_*) \ge 1 - \alpha$. By use of Sklar's Theorem \citep{sklar1973random} to express the c.d.f. in terms of a copula $C_{\mathbf{X}}$, and defining a vector of probabilities $(q_{*,1}, \dots, q_{*,m}) = (F_{X_1}(x_{*,1}), \dots, F_{X_m}(x_{*,m})) \in [0,1]^m$ restricted to a fixed value $q_{*} = q_{*,1} = \dots = q_{*,m}$, we can restate the problem as the search for a solution
\begin{equation}
    \min q_* \text{  subject to  } C_{\mathbf{X}}(q_*, \dots, q_*) \ge 1-\alpha,
\label{eq:copula-opt-problem-0}
\end{equation}
where the $\min$ is motivated by our efficiency goals to best match target coverage $(1-\alpha)$ (\autoref{remark:equiv-of-events}).

\textbf{Obtaining a solution via \m{max-rank} (\autoref{methods:max-rank-optimality}).} For a given solution to \autoref{eq:copula-opt-problem-0} the desired quantiles with FWER control are then computed as $\mathbf{x}_* = (F_{X_1}^{-1}(q_*), \dots, F_{X_m}^{-1}(q_*)) \in \mathbb{R}^m$. These equate the quantiles $\hat{Q}_{1-\hat{\alpha}_1}, \dots, \hat{Q}_{1-\hat{\alpha}_m}$ described in \autoref{algo:max-rank}. Leveraging operations in the rank space, we show that \m{max-rank} provides such a solution $q_{\max}$ to the problem. We state this in the following proposition:
\begin{proposition}
    The \m{max-rank} procedure (\autoref{algo:max-rank}) provides a solution $q_{\max}$ to the constrained problem in \autoref{eq:copula-opt-problem-0} with FWER control at level $\alpha$.
\label{prop:opt-max-rank}
\end{proposition}
Furthermore, the $\min$ implies that the associated quantiles $\mathbf{x}_{\max}$ are optimally tight in terms of coverage, under the condition that $q_{\max} = q_{\max,1} = \dots = q_{\max,m}$.

\begin{figure*}[t]
    \centering
    \includegraphics[width=0.7\linewidth]{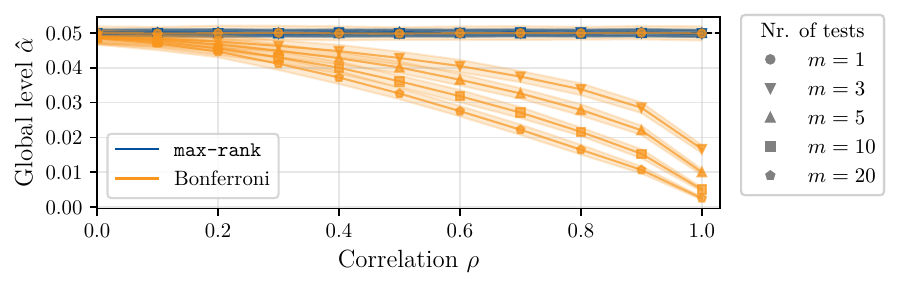}
    \vspace*{-2mm}
    \caption{
    We examine \emph{global} adjusted significance levels $\hat{\alpha}$ against the desired FWER control level $\alpha=0.05$ for varying correlations $\rho$ and number of tests $m$ (for fixed $n=100\,000$, shading denotes std. deviation across $100$ random trials). While Bonferroni becomes increasingly conservative as either correlation or test counts increase, \m{max-rank} provides stable and tight FWER control at target level.
    }
    \label{fig:global-level-corr}
    \vspace*{-2mm}
\end{figure*}

\begin{figure*}[t]
    \centering
    \includegraphics[width=0.7\linewidth]{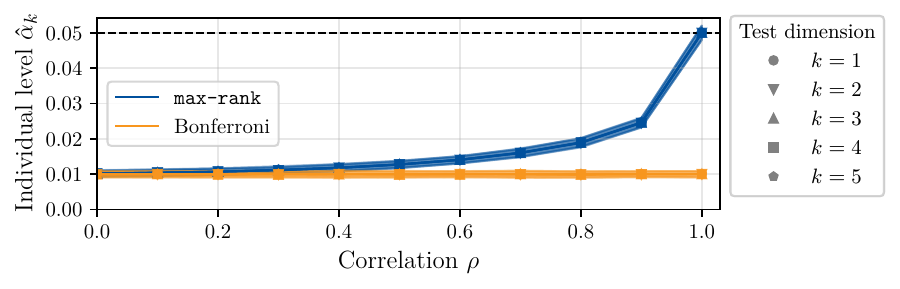}
    \vspace*{-2mm}
    \caption{
    We examine \emph{individual} adjusted significance levels $\hat{\alpha}_k$ for $k =1,\dots, 5$ against the ideal Type-I error control level $\alpha_k=0.05$ for varying correlations $\rho$ (for fixed $n=100\,000$, shading denotes std. deviation across $100$ random trials). While Bonferroni enforces highly conservative Type-I error levels to ensure FWER control, \m{max-rank} can gradually exploit positive test dependencies to improve its performance in each dimension.
    }
    \label{fig:indiv-level-corr}
    \vspace*{-2mm}
\end{figure*}

\textbf{Comparison to the Bonferroni correction (\autoref{methods:relation-to-bonferroni}).} Under independence the c.d.f. $F_{\mathbf{X}}(\mathbf{x}_*)$ factorizes, and the coverage constraint in \autoref{eq:copula-opt-problem-0} simplifies to ensuring $(q_{*})^m \ge 1-\alpha$. \m{max-rank} satisfies the condition directly by \autoref{prop:opt-max-rank}, \ie, we have $q^{\perp}_{\max} \geq (1-\alpha)^{1/m}$ denoting the solution under independence. In contrast, Bonferroni ensures the condition is met by $q_{\text{Bonf}} = (1 - \alpha/m)$, since $(1 - \alpha/m) \geq (1-\alpha)^{1/m}$ for any $m \geq 1,\,\alpha \in (0,1)$. It directly follows that $q^{\perp}_{\max} \leq q_{\text{Bonf}}$. In the case of positive dependency we impose that correlated tests follow \emph{positive lower orthant dependency} \citep{nelsen2006introduction}, suggesting a higher probability of taking on similar values than under independence. This directly implies that $q_{\max} \leq q^{\perp}_{\max}$. We summarize these results as:
\begin{proposition}
    The solution $q_{\max}$ provided by the \m{max-rank} procedure (\autoref{algo:max-rank}) to \autoref{eq:copula-opt-problem-0} ensures that we have $q_{\max} \leq q^{\perp}_{\max} \leq q_{\text{Bonf}}$ in the case of both independent and positively dependent test statistics.
\label{prop:better-than-bonf}
\end{proposition}

\section{SIMULATION STUDY}
\label{sec:experiments}

We empirically demonstrate our two results in \autoref{prop:opt-max-rank} and \autoref{prop:better-than-bonf} via a simulation study. We simulate $\mathbf{X}$ as a multivariate random variable whose $m$ components may be jointly positively correlated. Specifically, we simulate $\mathbf{X} \sim \mathcal{N}(\bm{\mu}_m, \bm{\Sigma}_{m\times m})$ with $\bm{\mu}_m = \mathbf{0}_m$ and $\bm{\Sigma}_{m \times m} = \rho\,\cdot\,\mathbf{1}_{m \times m} + (1-\rho)\,\cdot\,\mathbf{I}_{m \times m}$. Thus, covariance entries on the off-diagonals are determined by the correlation strength $\rho \in [0,1]$, while the diagonal denotes perfect correlation. The desired (global) significance level for FWER control is fixed at $\alpha=0.05$, and we vary the correlation strength $\rho$, number of tests $m$, and number of statistics $n$ for every component $X_k$ of $\mathbf{X}$. 

\textbf{Global FWER control.} We compare the adjusted significance levels produced by both \m{max-rank} and Bonferroni across varying correlation strengths and for $m \in \{1, 3, 5, 10, 20\}$ in \autoref{fig:global-level-corr}. We observe two key drawbacks for Bonferroni. Firstly, the correction becomes increasingly conservative as correlation between tests increases, with a FWER control level tending towards zero. Secondly, the rate accelerates as the number of tests increases. In contrast, \m{max-rank} suffers from neither of these behaviours and tightly maintains FWER control at target level $\alpha$, ensuring consistent results independently of correlation strength and test dimensionality (\autoref{prop:opt-max-rank}).

\textbf{Individual Type-I error control.} We next compare the adjusted significance levels obtained by \m{max-rank} and Bonferroni for \emph{individual} testing dimensions $k \in [m]$ and for fixed $m=5$ against varying correlation strengths in \autoref{fig:indiv-level-corr}. Bonferroni limits individual Type-I error levels to $\hat{\alpha}_{\text{Bonf}} = \alpha/m = 0.01$ in order to achieve FWER control, forcing overly conservative behaviour in each test. In contrast, \m{max-rank} is able to exploit any positive dependency between tests, resulting in reduced requirements on individual Type-I error levels as the correlation increases (\autoref{prop:better-than-bonf}). This, of course, strongly benefits the chances of a significant result in each dimension. Meanwhile, we observe in \autoref{fig:global-level-corr} for $m=5$ that the FWER remains controlled.

\textbf{FWER control and sample size (\autoref{subsec:app-exp-simulation}).} Since \m{max-rank} relates to \cite{westfall1993resampling} and is thus interpretable as a resampling-based correction, we anticipate the sample size $n$ (equivalent to the number of resamples) to affect FWER control, in particular when running many tests in parallel. Thus, we verify in \autoref{fig:global-level-test-samp} adjusted significance levels for \m{max-rank} and Bonferroni across varying combinations of correlation $\rho$, sample size $n$ and test counts $m$. A direct correspondence between FWER control and the magnitudes of $m$ and $n$ is apparent: as we run more tests in parallel the need for additional samples increases in order to empirically maintain FWER control at target level $\alpha$ (\eg, roughly $n \geq 100\,m$ for $\rho=0$). Crucially, larger correlations suggest a lower need for additional samples as \m{max-rank} can exploit existing rank order information, whereas Bonferroni merely grows more conservative.

\section{APPLICATION TO CONFORMAL PREDICTION}
\label{sec:conformal}

We further elaborate on the usefulness of our proposed \m{max-rank} correction for conformal prediction. First, we illustrate how multiple testing issues emerge when running multiple CP procedures in parallel. We then theoretically motivate why \m{max-rank} integrates well with the conformal design and its underlying conditions, and empirically validate its effectiveness in two CP applications that require a multiple testing correction.

\textbf{Multiple testing issues in conformal prediction.} We previously discussed that running multiple conformal procedures at once equates permutation testing in parallel at a fixed significance level $\alpha$. By \autoref{eq:equiv-of-events} and \autoref{remark:control-as-coverage} this is identical to ensuring a $(1-\alpha)$ coverage rate for each component of an $m$-dimensional response $\mathbf{Y}_{n+1} = (Y_{n+1, 1}, \dots, Y_{n+1, m})$. The same intuition is employed in our theoretical formulation for \autoref{eq:copula-opt-problem-0}. Take $\mathbf{X}_{n+1} = (X_{n+1, 1}, \dots, X_{n+1, m})$ to be a matching multivariate input, then we can directly observe how the coverage guarantee in \autoref{eq:conf-guarant} can be violated since
\begin{alignat}{1}
    \mathbb{P}\Big( & \mathbf{Y}_{n+1} \in \hat{C}(\mathbf{X}_{n+1}) \Big) \nonumber \\
    &= \mathbb{P} \Big( \bigcap_{k=1}^{m} \big( Y_{n+1, k} \in \hat{C}(X_{n+1, k}) \big) \Big) \nonumber \\ 
    &= 1 - \mathbb{P} \Big( \bigcup_{k=1}^{m} \big( Y_{n+1, k} \notin \hat{C}(X_{n+1, k}) \big) \Big) \nonumber \\
    &\ge 1 - \sum_{k=1}^{m} \mathbb{P} \Big( Y_{n+1, k} \notin \hat{C}(X_{n+1, k}) \Big) \nonumber \\
    &= 1 - \sum_{k=1}^{m} \mathbb{P} \Big( \hat{P}_{n+1, k}(Y_{n+1, k}; S) \le \alpha_k \Big) \nonumber \\
    &\ge 1 - \sum_{k=1}^{m} \alpha_k = 1 - m \alpha,  
    \quad \text{for } \alpha_k=\alpha \text{ fixed.} 
    \label{eq:mht-problem}
\end{alignat}
That is, guaranteed coverage is limited to $1-m\alpha \le 1-\alpha$ for any number of tests $m \in \mathbb{N}^+$, requiring a correction. 

A key observation is that due to their joint dependence on common calibration data $\mathcal{D}_{cal}$, obtained conformal p-values in any single dimension are bound to exhibit some degree of positive dependency, a property known as \emph{positive regression dependent on a subset} (PRDS) \citep{y.benjamini2001, s.bates2022, z.chi2022}. This highlights that any MHT correction relying on an independence assumption (such as Bonferroni) is bound to be overly conservative \emph{a priori}, and our empirical results point towards a similar positive dependency \emph{across} testing dimensions for CP settings.

\subsection{Using \m{max-rank} for conformal prediction}
\label{subsec:conf-max-rank-0}

Since we have closely connected \m{max-rank}'s formulation to conformal prediction, one can rightly assume that it integrates well with any conformal procedure following \autoref{algo:split-cp}. To rigorously assert this, we verify that \m{max-rank} meets two key requirements of conformal prediction: \emph{(i)} the exchangeability condition on the underlying data is preserved, and \emph{(ii)} \autoref{eq:mht-problem} is corrected in such a way that the conformal coverage guarantee from \autoref{eq:conf-guarant} is upheld.

We demonstrate in \autoref{subsec:conf-max-rank} that \m{max-rank} satisfies both conditions by analysing its two key operations, namely \emph{(i)} operating in the rank space of nonconformity scores, and \emph{(ii)} applying a $\max$-operation on these ranks (or more generally, any set of exchangeable random variables). While the requirement on valid coverage technically directly follows from \m{max-rank}'s FWER-controlling property (\autoref{prop:opt-max-rank}), an additional proof is provided via rank coverage. We summarize these insights in the following proposition:
\begin{proposition}
    The \m{max-rank} procedure (\autoref{algo:max-rank}) provides a multiple testing correction for conformal prediction that preserves exchangeability and ensures valid coverage (\autoref{eq:conf-guarant}).
\label{prop:max-rank-for-conf}
\end{proposition}

\vspace{-2mm}
\subsection{Conformal prediction experiments} 
\label{subsec:conf-exp-reg}
\vspace{-1mm}

\begin{figure*}[t]
    \centering
    % First image in a minipage
    \begin{minipage}[t]{0.49\linewidth}
        \centering
        \includegraphics[width=\linewidth]{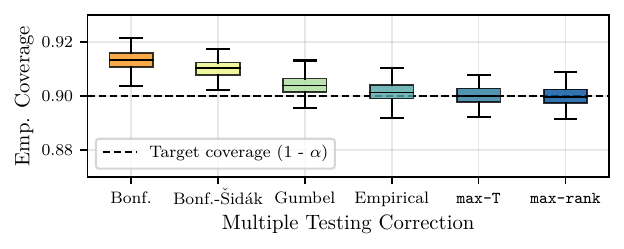}
    \end{minipage}%
    \hfill
    % Second image in a minipage
    \begin{minipage}[t]{0.49\linewidth}
        \centering
        \includegraphics[width=\linewidth]{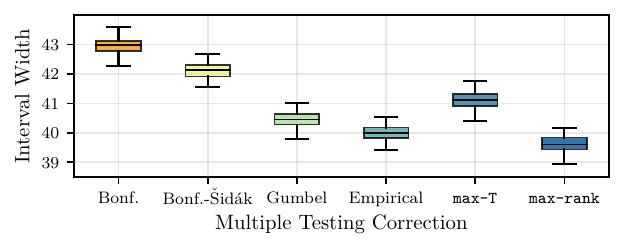}
    \end{minipage}
    \caption{
    Empirical coverage (\emph{left}) and mean prediction interval width (\emph{right}) for different corrections ($\alpha=0.1$) on BDD100k. Results are averaged across objects from multiple classes (see \autoref{subsec:app-exp-objdetect}), while boxplots depict the distribution over $100$ random trials. Inference times in sec. per trial (\emph{left to right} by method): $2.7$, $2.7$, $17.8$, $62.6$, $2.7$, and $2.95$. \m{max-rank} outperforms other corrections with notably lower runtimes than copula-based methods.
    }
    \label{fig:bdd100k-res}
\end{figure*}

\begin{table*}[t]
    \centering
    \small
    \setlength{\tabcolsep}{6pt}
    \begin{tabular}{lcccccc}
         \toprule
         \multicolumn{1}{c}{} & \multicolumn{2}{c}{\textbf{scpf ($m=3$)}} & \multicolumn{2}{c}{\textbf{rf1 ($m=8$)}} & \multicolumn{2}{c}{\textbf{scm1d ($m=16$)}} \\
         \cmidrule(lr){2-3} \cmidrule(lr){4-5} \cmidrule(lr){6-7}
         \textbf{Correction} & Coverage & Interval Width & Coverage & Interval Width & Coverage & Interval Width \\
         \toprule
         No correction & $0.83 \pm 0.04$ & $18.89 \pm 3.62$ & $0.50 \pm 0.01$ & $1.88 \pm 0.05$ & $0.50 \pm 0.02$ & $243.87 \pm 4.50$ \\
         Bonferroni & $0.95 \pm 0.02$ & $46.73 \pm 13.39$ & $0.92 \pm 0.01$ & $7.18 \pm 0.44$ & $0.96 \pm 0.01$ & $793.70 \pm 50.65$ \\
         Bonf.-\v{S}id\'{a}k & $0.95 \pm 0.02$ & $46.73 \pm 13.39$ & $0.92 \pm 0.01$ & $6.98 \pm 0.43$ & $0.96 \pm 0.01$ & $793.70 \pm 50.65$ \\
         Gumbel Copula & $0.90 \pm 0.03$ & $\boldsymbol{30.15 \pm 5.53}$ & $0.91 \pm 0.01$ & $6.55 \pm 0.39$ & $0.94 \pm 0.01$ & $697.89 \pm 36.48$ \\
         Emp. Copula & $0.92 \pm 0.03$ & $34.38 \pm 6.74$ & $0.91 \pm 0.01$ & $\underline{6.32 \pm 0.31}$ & $0.91 \pm 0.01$ & $\underline{579.21 \pm 15.43}$ \\
         \m{max-T} & $0.91 \pm 0.03$ & $53.77 \pm 10.63$ & $0.90 \pm 0.01$ & $6.76 \pm 0.24$ & $0.90 \pm 0.01$ & $581.93 \pm 14.14$ \\
         \m{max-rank} (Ours) & $0.91 \pm 0.03$ & $\underline{32.28 \pm 6.26}$ & $0.90 \pm 0.01$ & $\boldsymbol{6.08 \pm 0.29}$ & $0.90 \pm 0.01$ & $\boldsymbol{564.01 \pm 13.79}$ \\
         \bottomrule
    \end{tabular}
    \caption{
    Empirical coverage and mean prediction interval width for different corrections ($\alpha=0.1$) on three multi-target datasets: scpf (1137 samples), rf1 (9125 samples), and scm1d (9803 samples). We employ a random forest regressor \citep{scikit-learn} and $20\%$ calibration split, and report means and std. deviation across $100$ random trials. \m{max-rank} consistently provides strong results (\textbf{best}, \underline{second best} results).
    } 
    \label{tab:conf-reg}
\end{table*}

Finally, we showcase \m{max-rank}'s usefulness experimentally for two applications, conformal multi-target regression and conformal object detection. Our code is available at \url{https://github.com/alextimans/max-rank}.

\textbf{Experiment design.} We consider applying conformal prediction to multi-target regression problems where $m$ real-valued and possibly dependent targets are predicted simultaneously. We aim to ensure that all $m$ targets are jointly covered with probability $(1-\alpha)$ by our conformal procedure, and consider a range of multiple testing corrections for that purpose: Bonferroni and its slightly more powerful \v{S}id\'{a}k-adjustment \citep{abdi2007sidak}, two copula-based corrections from \cite{messoudi2021copula}, \m{max-T} \citep{westfall1993resampling} and finally \m{max-rank}. We include the copula-based corrections because they also address settings of positive dependency, and since \m{max-rank}'s theoretical motivation via copulas provides an interesting link. For CP (\autoref{algo:split-cp}) we employ nonconformity scores of the simple residual form $s_{i,k} = |Y_{i,k} - \hat{f}(X_{i,k})|$ where $i \in [n], k \in [m]$, and construct the prediction intervals for test points as $\hat{C}(X_{n+1,k})=[\hat{f}(X_{n+1,k}) - \hat{Q}_{1-\hat{\alpha}_k},\,\hat{f}(X_{n+1,k}) + \hat{Q}_{1-\hat{\alpha}_k}]$, where $\hat{Q}_{1-\hat{\alpha}_k}$ again denotes the conformal quantile at \emph{adjusted} individual significance level $\hat{\alpha}_k$ using a MHT correction. We compare results with different corrections across two key conformal metrics \citep{v.vovk2005}: empirical coverage as a measure of \emph{validity} to verify \autoref{eq:conf-guarant}, and average prediction interval width to assess the \emph{efficiency} of the solution. See also \autoref{app:experiments}. 

\textbf{Conformal multi-target regression.} Following \cite{messoudi2021copula} we consider three tabular datasets with a varying number of targets\footnote{Obtainable at \url{https://github.com/tsoumakas/mulan}}: scpf (website traffic data with $m=3$), rf1 (river flow data with $m=8$), and scm1d (product pricing data with $m=16$). Our results for the usual target coverage of $(1-\alpha)=90\%$ are displayed in \autoref{tab:conf-reg}. Firstly, we observe that all corrections provide valid coverage, whereas without we strongly undercover. Among corrections, Bonferroni and Bonferroni-Šidák provide overly conservative results, and copula methods perform generally well. In particular for the scpf dataset the parametric Gumbel copula benefits from the low test dimensionality ($m=3$), while \m{max-T} and \m{max-rank} suffer from its small sample size. Overall, \m{max-rank} tends to outperform other corrections in terms of interval width, and matches the target coverage level very closely. 

\textbf{Conformal object detection.} We also explore the more complex task of predicting an object's location in an image via its bounding box coordinates ($m=4$) with a deep learning-based object detector \citep{wu2019detectron2}. Following \cite{mukama2024copula} we use the large-scale BDD100k dataset \citep{yu2020bdd}, but apply CP separately to multiple classes (as opposed to only considering a single class `car')\footnote{In concurrent work, we develop a richer conformal methodology for uncertainty in object detection and use \m{max-rank} as a correction \citep{timans2024conformalod}.}. Our results in \autoref{fig:bdd100k-res} stress that \m{max-rank} ensures both tight empirical coverage and the smallest prediction intervals across considered methods. Copula-based corrections provide competitive results but suffer from substantially larger computational requirements. The Gumbel copula requires numerical optimization to determine its parametric form, whereas the empirical copula repeatedly evaluates ranks across all $m$ dimensions during inference. In contrast, \m{max-rank} collapses the test dimensions and its simple quantile operations result in low runtimes comparable to Bonferroni and \m{max-T}.

\section{DISCUSSION}
\label{sec:conclusion}

We introduce \m{max-rank}, an efficient MHT correction for testing under positive dependency which adapts \cite{westfall1993resampling}'s corrections to conformal prediction. While we provide strong theoretical and empirical results, our method's benefits are mostly restricted to \emph{(i)} data dependencies that become apparent through rank ordering, \emph{(ii)} settings of positive dependency, since we do not benefit from rank information under negative dependency and thus do not improve upon Bonferroni, and \emph{(iii)} its permutation-based, data-driven nature necessitates a sufficient ratio of test statistics $n$ to test dimensions $m$ for empirical FWER control. Future work includes establishing stronger theoretical results, such as on convergence rates to target level $\alpha$, the dependence between quantities $m$, $n$, and $\alpha$, and its relationship to the empirical copula. Naturally, we may also experimentally extend to further applications such as multi-step time series \citep{sun2023copula}. We hope that \m{max-rank} contributes to the ongoing research on multiple testing under dependency, all the while addressing practical needs in predictive settings using conformal prediction.

%%%%%%%%% ADDITIONAL
\subsubsection*{Acknowledgements}
We thank Aaditya Ramdas and Jelle Goeman for making us aware of connections to the Westfall \& Young testing procedures, and Drew Prinster for helpful feedback on an earlier draft. This project was generously supported by the Bosch Center for Artificial Intelligence. Eric Nalisnick did not use any resources from Johns Hopkins for this work. \textbf{Author contributions are as follows}: Alexander Timans established the theoretical connections to conformal prediction and Westfall \& Young, performed experiments and paper writing. Christoph-Nikolas Straehle developed the \m{max-rank} algorithm and proved several theoretical properties. Kaspar Sakmann, Christian A. Naesseth and Eric Nalisnick provided project guidance and helpful feedback.

%%%%%%%%% REFERENCES
\bibliography{paper}
\bibliographystyle{plainnat}

%%%%%%%%% CHECKLIST
\clearpage
\onecolumn

\section*{CHECKLIST}

 \begin{enumerate}

 \item For all models and algorithms presented, check if you include:
 \begin{enumerate}
   \item A clear description of the mathematical setting, assumptions, algorithm, and/or model. [\yes{Yes}] Algorithmic descriptions are included in the paper, the full proofs are included in the appendix.
   \item An analysis of the properties and complexity (time, space, sample size) of any algorithm. [\yes{Yes}] We discuss the algorithmic properties and empirically verify sample size and runtime requirements, but do not include a formal theoretical analysis. 
   \item (Optional) Anonymized source code, with specification of all dependencies, including external libraries. [\yes{Yes}] Our code is publicly available at \url{https://github.com/alextimans/max-rank}.
 \end{enumerate}

 \item For any theoretical claim, check if you include:
 \begin{enumerate}
   \item Statements of the full set of assumptions of all theoretical results. [\yes{Yes}] The full proofs are included in the appendix.
   \item Complete proofs of all theoretical results. [\yes{Yes}] The full proofs are included in the appendix.
   \item Clear explanations of any assumptions. [\yes{Yes}] The full proofs are included in the appendix.
 \end{enumerate}

 \item For all figures and tables that present empirical results, check if you include:
 \begin{enumerate}
   \item The code, data, and instructions needed to reproduce the main experimental results (either in the supplemental material or as a URL). [\yes{Yes}] We provide experimental design details in the paper and the appendix, and the code is publicly available.
   \item All the training details (e.g., data splits, hyperparameters, how they were chosen). [\yes{Yes}]
   \item A clear definition of the specific measure or statistics and error bars (e.g., with respect to the random seed after running experiments multiple times). [\yes{Yes}]
   \item A description of the computing infrastructure used. (e.g., type of GPUs, internal cluster, or cloud provider). [\yes{Yes}] All results are reproducible on a single consumer-grade GPU (\eg, NVIDIA RTX 4090). \m{max-rank} is computationally light-weight and is easily runnable on CPU for smaller experiments.
 \end{enumerate}

 \item If you are using existing assets (e.g., code, data, models) or curating/releasing new assets, check if you include:
 \begin{enumerate}
   \item Citations of the creator If your work uses existing assets. [\yes{Yes}]
   \item The license information of the assets, if applicable. [\yes{Yes}] Our code is publicly released under an AGPL-3.0 license.
   \item New assets either in the supplemental material or as a URL, if applicable. [\yes{Yes}] The code is available via an URL.
   \item Information about consent from data providers/curators. [\na{Not Applicable}] Used datasets are synthetic or publicly available.
   \item Discussion of sensible content if applicable, e.g., personally identifiable information or offensive content. [\na{Not Applicable}]
 \end{enumerate}

 \item If you used crowdsourcing or conducted research with human subjects, check if you include:
 \begin{enumerate}
   \item The full text of instructions given to participants and screenshots. [\na{Not Applicable}]
   \item Descriptions of potential participant risks, with links to Institutional Review Board (IRB) approvals if applicable. [\na{Not Applicable}]
   \item The estimated hourly wage paid to participants and the total amount spent on participant compensation. [\na{Not Applicable}]
 \end{enumerate}

 \end{enumerate}

%%%%%%%%% APPENDIX
\clearpage
\appendix
\onecolumn

\aistatsappendixtitle{Max-Rank: Efficient Multiple Testing for Conformal Prediction \\-- Supplementary Materials --}
% \begin{center}
%     {\Large \textbf{Max-Rank: Efficient Multiple Testing for Conformal Prediction}} \\
%     {\Large \textbf{-- Supplementary Materials --}}
% \end{center}
\label{appendix}

The appendix is organized as follows:
\begin{itemize}
    \item Mathematical details and proofs in \autoref{app:math}
    \item Algorithmic details on \m{max-T} and \m{min-P} in \autoref{app:algo}
    \item Additional background on FDR control and the Bonferroni correction in \autoref{app:background}
    \item Additional experiment details for our simulation study, multiple testing baselines, metrics, and conformal experiments in \autoref{app:experiments}
\end{itemize}

\begin{table}[!h]
    \centering
    \small
    \setlength{\tabcolsep}{6pt}
    \resizebox{\textwidth}{!}{
    \begin{tabular}{ccl}
         \toprule
         \textbf{Notation} & \textbf{Type} & \textbf{Description} \\
         \midrule
         $n$ & $\in \mathbb{N}^+$ & Nr. of calibration samples, nr. of test statistics, index $i=1, \dots, n$ \\
         
         $m$ & $\in \mathbb{N}^+$ & Nr. of conformal procedures, nr. of test dimensions, index $k=1, \dots, m$ \\
         
         $\alpha, \alpha_k$ & $\in (0, 1)$ & Target miscoverage rate, target significance level (global, in the $k$-th dimension) \\
         
         $\hat{\alpha}, \hat{\alpha}_k$ & $\in (0, 1)$ & Adjusted or corrected significance level (global, in the $k$-th dimension) \\
         
         $\hat{f}(\cdot)$ & $\mathcal{X} \rightarrow \mathcal{Y}$ & Pre-trained prediction model \\
         
         $s(\cdot)$ & $\mathcal{X} \times \mathcal{Y} \rightarrow \mathbb{R}$ & Conformal scoring function \\
         
         $s_i, s_{i,k}$ & $\in \mathbb{R}$ & Nonconformity score for the $i$-th sample, test statistic (in the $k$-th dimension) \\
         
         $r_i, r_{i,k}$ & $\in \mathbb{N}$ & Rank of nonconformity score for the $i$-th sample (in the $k$-th dimension) \\
         
         $S$ & $\subset \mathbb{R}$ & Set of scores $\{s_1, \dots, s_n\}$ given a single dimension (equiv. to fixed $\mathbf{s}_k$ ignoring ordering) \\
         
         $\hat{C}(X_{n+1})$ & $\subseteq \mathcal{Y}$ & Conformal prediction set for test input $X_{n+1}$ \\
         
         $\hat{P}_{n+1}(y; S)$ & $\in [0,1]$ & Conformal p-value for the null hypothesis $H_0: Y_{n+1} = y$ based on $S$ \\
         
         $\hat{Q}(1-\alpha; S),\,\hat{Q}_{1-\alpha}$ & $\in \mathbb{R}$ & Conformal quantile at target coverage $(1-\alpha)$ based on $S$ \\

         $\hat{Q}(1 - \alpha_k; \mathbf{s}_k), \hat{Q}_{1 - \alpha_k}$ & $\in \mathbb{R}$ & Conformal quantile at target coverage $(1 - \alpha_k)$ based on $\mathbf{s}_k$ (\ie, in the $k$-th dimension) \\
         
         $\mathbf{s}_k$ & $\in \mathbb{R}^n$ & Vector of scores, test statistics, empirical null distribution (in the $k$-th dimension) \\
         
         $\mathbf{r}_k$ & $\in \mathbb{N}^n$ & Vector of score ranks, empirical null distribution in rank space (in the $k$-th dimension) \\
         
         $\mathbf{S}$ & $\in \mathbb{R}^{n \times m}$ & Matrix of test statistics with columns $\mathbf{s}_1, \dots, \mathbf{s}_m$ \\
         
         $\mathbf{R}$ & $\in \mathbb{N}^{n \times m}$ & Matrix of column-wise ranks of $\mathbf{S}$ \\
         
         $\mathbf{r}_{\max}$ & $\in \mathbb{N}^n$ & Vector of max. score ranks, composite empirical null distribution in rank space \\

         $r_{\max}$ & $\in \mathbb{N}$ & The rank value $\hat{Q}(1 - \alpha; \mathbf{r}_{\max})$ representing the corrected significance levels $\hat{\alpha}, \hat{\alpha}_k$ \\
         
         \bottomrule
    \end{tabular}
    }
    \caption{
    Notation table with clarifications for key quantities presented in the main paper.
    } 
    \label{tab:notation}
\end{table}

\section{MATHEMATICAL DETAILS}
\label{app:math}

This section includes notation clarification in \autoref{tab:notation}, and the details of our theoretical results for \m{max-rank} (\autoref{prop:opt-max-rank} and \autoref{prop:better-than-bonf} in \autoref{sec:theory}), as well as its use for conformal prediction (\autoref{prop:max-rank-for-conf} in \autoref{subsec:conf-max-rank-0}).

\subsection{Recasting the multiple testing problem}
\label{subsec:methods-copula-formulation}

We begin by reformulating our multiple testing problem in terms of multivariate copulas. We show that the search for an adjusted global significance level with FWER control can be recast as the search for a probability vector satisfying a coverage constraint under a multivariate copula.

Let $\mathbf{X} = (X_1, \dots, X_m) \in \mathbb{R}^m$ be an $m$-dimensional random variable where each dimension $k \in [m]$ corresponds to a hypothesis test. We denote the multivariate cumulative distribution function (c.d.f.) of $\mathbf{X}$ as $F_{\mathbf{X}}(\mathbf{x}) = \mathbb{P}(X_1 \le x_{1}, \dots, X_m \le x_{m})$, for any realization $\mathbf{x}=(x_1, \dots, x_m)$. As such $F_{\mathbf{X}}(\mathbf{x})$ is right-continuous and non-decreasing in its arguments. Looking at the FWER control definition in \autoref{eq:fwer} we see that $\mathbb{P}(m_{\text{FP}} \geq 1) = 1 - \mathbb{P}(m_{\text{FP}} = 0) \leq \alpha \Leftrightarrow \mathbb{P}(m_{\text{FP}} = 0) \geq 1 - \alpha$, \ie, the probability of not producing \emph{any} false positives or Type-I errors across test dimensions is kept above $(1-\alpha)$. Since by \autoref{remark:control-as-coverage} such a Type-I error control in any \emph{single} test dimension can be expressed in terms of a quantile $x_k$ ensuring $(1-\alpha)$ coverage, a \emph{joint} Type-I error control across tests matching the FWER is ensured for any quantile vector $\mathbf{x_*}=(x_{*,1}, \dots, x_{*,m})$ satisfying
\begin{equation}
    F_{\mathbf{X}}(\mathbf{x}_*) \ge 1 - \alpha.
\label{eq:cdf-threshold}
\end{equation}
Such an instance $\mathbf{x_*}$ always exists since $F_{\mathbf{X}}(\mathbf{x_*}) \rightarrow 0$ if any $x_{*,k} \rightarrow -\infty$ and $F_{\mathbf{X}}(\mathbf{x_*}) \rightarrow 1$ as all $x_{*,k} \rightarrow \infty$. 

Alternatively, we can use Sklar's Theorem \citep{sklar1973random} to express the c.d.f. $F_{\mathbf{X}}(\mathbf{x})$ in terms of a multivariate copula $C_{\mathbf{X}}(\mathbf{u})$ with uniform marginals $\mathbf{u} \in [0,1]^m$ capturing the dependency structure among tests. That is, we can express $F_{\mathbf{X}}(\mathbf{x_*})$ as $C_{\mathbf{X}}(F_{X_1}(x_{*,1}), \dots, F_{X_m}(x_{*,m}))$, and if we explicitly define a vector of probabilities $\mathbf{q_*} = (q_{*,1}, \dots, q_{*,m}) = (F_{X_1}(x_{*,1}), \dots, F_{X_m}(x_{*,m})) \in [0,1]^m$, as $C_{\mathbf{X}}(q_{*,1}, \dots, q_{*,m})$. Thus the coverage target from \autoref{eq:cdf-threshold} can be re-stated as
\begin{equation}
    F_{\mathbf{X}}(\mathbf{x_*}) = \mathbb{P}(X_1 \le F_{X_1}^{-1}(q_{*,1}), \dots, X_m \le F_{X_m}^{-1}(q_{*,m})) = C_{\mathbf{X}}(q_{*,1}, \dots, q_{*,m}) \ge 1 - \alpha.
\label{eq:copula-threshold}
\end{equation}
Since \m{max-rank} computes a single rank quantile which informs the conformal quantile in \emph{each} test dimension, we restrict $\mathbf{q}_*$ to a single fixed quantile probability $q_{*} = q_{*,1} = \dots = q_{*,m}$, \ie, $\mathbf{q_*} = (q_*, \dots, q_*)$. The same restriction is made independently in \cite{messoudi2021copula} to simplify copula calculations. Motivated by our efficiency goals to match target coverage $(1-\alpha)$ as closely as possible and obtain the smallest conformal prediction sets (\autoref{remark:equiv-of-events}), we seek the smallest fixed probability $q_*$ satisfying \autoref{eq:copula-threshold}. Thus a solution to our initial FWER control problem can be restated as the search for 
\begin{equation}
    \min q_* \text{  subject to  } C_{\mathbf{X}}(q_*, \dots, q_*) \ge 1-\alpha.
\label{eq:copula-opt-problem}
\end{equation}
Such an instance $\mathbf{q_*} = (q_*, \dots, q_*)$ also always exists since $C_{\mathbf{X}}(\mathbf{q_*}) \rightarrow 0$ as $q_{*} \rightarrow -\infty$ and $C_{\mathbf{X}}(\mathbf{q_*}) \rightarrow 1$ as $q_{*} \rightarrow \infty$.

\subsection{Obtaining a solution via \m{max-rank}}
\label{methods:max-rank-optimality}

For a given solution $q_*$ satisfying \autoref{eq:copula-opt-problem}, the desired quantile vector ensuring FWER control is then computed by evaluating the probability under each inverse marginal as $\mathbf{x_*} = (F_{X_1}^{-1}(q_*), \dots, F_{X_m}^{-1}(q_*)) \in \mathbb{R}^m$. Note that now the elements of $\mathbf{x_*}$ may differ from each other, since the marginals are not necessarily identical. The obtained quantiles in $\mathbf{x_*}$ are equal to the adjusted quantiles $\hat{Q}_{1-\hat{\alpha}_1}, \dots, \hat{Q}_{1-\hat{\alpha}_m}$ described in \autoref{algo:max-rank}. By relating the element-wise ranks of $x_{*,k}$ under $\mathbf{S}$ to the marginals $F_{X_k}$ and establishing a correspondence between the copula and a joint c.d.f. in rank space, we next show that our \m{max-rank} correction provides a solution $q_{\max}$ to \autoref{eq:copula-opt-problem}. 

For that, we relate the algorithmic steps in \autoref{algo:max-rank} to our problem setting in \autoref{subsec:methods-copula-formulation}. Consider the matrix of test statistics $\mathbf{S} \in \mathbb{R}^{n \times m}$ first introduced in \autoref{sec:methods}, and denote by $S_k$ its $k$-th column corresponding to the $k$-th empirical null (in set notation, rather than as a vector $\mathbf{s}_k$). We then associate $S_k$ with the $k$-th dimension of $\mathbf{X}$, \ie, the elements $s_{i,k} \in S_k,\,i \in [n]$ match with the set of $n$ realizations $\{x_{1,k},\dots,x_{n,k}\}$ of $X_k$, and $S_k$ thus provides an empirical estimate of the marginal $F_{X_k}$. Next, we evaluate the element-wise ranks of $\mathbf{x_*}$ under $\mathbf{S}$. For any element $x_{*,k} \in \mathbf{x_*}$ a direct correspondence between its rank under $S_k$ (see \autoref{eq:rank-definition}) and marginal $F_{X_k}$ is given by
\begin{equation}
    \text{rank}(x_{*,k}; S_k) = r_{*,k} = n \cdot q_* = n \cdot F_{X_k}(x_{*,k}),
\label{eq:rank-scaling}
\end{equation}
that is, the rank of $x_{*,k}$ is precisely the rank that ensures $q_*$ probability mass below it. In particular, we then have that $q_* = r_{*,k}/n = \Tilde{r}_{*,k}$ for any $k \in [m]$, where $\Tilde{r}_{*,k} \in [0,1]$ defines the normalized rank amenable to a copula. Since we have distinct scores (\eg, by introducing jitter noise) and thus distinct ranks $\{1,\dots,n\}$, it follows that the same rank $\Tilde{r}_{*} = \Tilde{r}_{*,1} = \dots = \Tilde{r}_{*,m}$ is retrieved in every dimension. Thus, we can establish an equivalence between the copula $C_{\mathbf{X}}$ which describes the joint c.d.f. of $\mathbf{X}$, and the matching joint c.d.f. in the (normalized) \emph{rank space} of $\mathbf{X}$, which we denote $F_{\Tilde{\mathbf{R}}}$. We use $\Tilde{\mathbf{R}}$ to stress the association with the matrix of rank statistics $\mathbf{R} \in \mathbb{N}^{n \times m}$ described in \autoref{sec:methods}. Leveraging a $\max$-operation, we can then state that
\begin{equation}
    C_{\mathbf{X}}(q_*,\dots,q_*) = F_{\Tilde{\mathbf{R}}}(\Tilde{r}_*, \dots, \Tilde{r}_*) = \mathbb{P}(\Tilde{R}_1 \leq \Tilde{r}_*, \dots, \Tilde{R}_m \leq \Tilde{r}_*) = \mathbb{P}(\max_{1 \leq k \leq m} \Tilde{R}_k \leq \Tilde{r}_*) = F_{\max \mathbf{\Tilde{R}}}(\Tilde{r}_*),
\label{eq:copula-max-rank}
\end{equation}
where $F_{\max \mathbf{\Tilde{R}}}$ now denotes the one-dimensional c.d.f. in the space of \emph{maximum} ranks across dimensions $k \in [m]$. Intuitively, ensuring that the condition holds jointly for ranks in all dimensions is equivalent to satisfying it in the `worst' one. $F_{\max \mathbf{\Tilde{R}}}$ is precisely the space in which \m{max-rank} operates after the $l^{\infty}$-norm is applied (matching the composite null $\mathbf{r}_{\max}$ in the paper), and in which following \autoref{algo:max-rank} a rank $\Tilde{r}_{\max} = r_{\max}/n = q_{\max}$ with probability mass $(1-\alpha)$ below it is computed. Thus we can employ \autoref{eq:copula-max-rank} in its reverse direction to claim 
\begin{equation}
    1 - \alpha \leq F_{\max \mathbf{\Tilde{R}}}(\Tilde{r}_{\max}) = F_{\mathbf{\Tilde{R}}}(\Tilde{r}_{\max}, \dots, \Tilde{r}_{\max}) = C_{\mathbf{X}}(q_{\max}, \dots, q_{\max}),
\label{eq:copula-max-rank-solution}
\end{equation}
therefore providing a solution which satisfies the constraint in \autoref{eq:copula-opt-problem}. Furthermore, since we precisely compute the rank at target level $(1-\alpha)$ (and ignoring any finite-sample correction), our solution $q_{\max}$ also satisfies the $\min$ in \autoref{eq:copula-opt-problem}, and the obtained quantiles $\mathbf{x}_{\max}$ are optimally tight in terms of coverage under the imposed restriction that $q_{\max} = q_{\max,1} = \dots = q_{\max,m}$. Note that if the restriction is lifted, marginally tighter individual quantiles $x_{\max,k}$ are potentially obtainable in some dimensions. However, it becomes unclear how the required algorithmic modifications to \m{max-rank} would continue to formally uphold FWER control.

We summarize our arguments by restating our proposition from \autoref{sec:theory}, which follows from the above:
\begin{proposition1}
    The \m{max-rank} procedure (\autoref{algo:max-rank}) provides a solution $q_{\max}$ to the constrained problem in \autoref{eq:copula-opt-problem-0} with FWER control at level $\alpha$.
% \label{prop:opt-max-rank}
\end{proposition1}

\subsection{Comparison to the Bonferroni correction}
\label{methods:relation-to-bonferroni}

Finally we compare our \m{max-rank} procedure to the Bonferroni correction, and demonstrate that \m{max-rank} performs either on par or better than Bonferroni under settings of independence or positive dependency between hypothesis tests. We first consider the case of independence, which the Bonferroni correction generally assumes.

\textbf{Under independence.} If the $m$ dimensions of $\mathbf{X}$ are considered independent the joint c.d.f. $F_{\mathbf{X}}(\mathbf{x})$ factorizes, and we then have by \autoref{eq:copula-threshold} that the coverage constraint in \autoref{eq:copula-opt-problem} reduces to
\begin{equation}
    C^{\perp}_{\mathbf{X}}(q_*, \dots, q_*) = F_{\mathbf{X}}(\mathbf{x}_*) = \prod_{k=1}^m F_{X_k}(x_{*,k}) = \prod_{k=1}^m F_{X_k}(F_{X_k}^{-1}(q_{*})) = \prod_{k=1}^m q_* = (q_{*})^m \ge 1-\alpha,
\label{eq:cdf-indep}
\end{equation}
where we refer to $C^{\perp}_{\mathbf{X}}$ as the factorized `independent copula' \citep{sklar1973random, messoudi2021copula}. \m{max-rank} solves the constraint by \autoref{prop:opt-max-rank} and is optimally tight, thus the solution under independence $q^{\perp}_{\max}$ directly satisfies $q^{\perp}_{\max} \geq (1-\alpha)^{1/m}$. In contrast, the Bonferroni correction ensures the condition is met by $q_{\text{Bonf}} = (1 - \alpha/m)$, since $(1 - \alpha/m) \geq (1-\alpha)^{1/m}$ for any $m \geq 1,\,\alpha \in (0,1)$. It directly follows that $q^{\perp}_{\max} \leq q_{\text{Bonf}}$, and \m{max-rank} therefore performs at least as well as the Bonferroni correction under independence.

\textbf{Under positive dependency.} In this setting we assume that the components $X_1, \dots, X_m$ of $\mathbf{X}$ exhibit a positive correlation known as \emph{positive lower orthant dependency}, which is defined as follows:
\begin{definition}[Positive lower orthant dependency (PLOD), \citep{nelsen2006introduction}, Def. 5.7.1]
    An $m$-dimensional random vector $\mathbf{X} = (X_1, \dots, X_m)$ is PLOD if for all $\mathbf{x} = (x_1, \dots, x_m) \in \mathbb{R}^m$ we have that
    $$\mathbb{P}(\mathbf{X} \leq \mathbf{x}) \geq \prod_{k=1}^m \mathbb{P}(X_k \leq x_k).$$
\label{def:plod}
\end{definition}
\vspace{-5mm}
That is, the condition supposes that the probability of components jointly taking on large (or small) values is at least as large as if all components were considered independent. \emph{PLOD} provides a sensible notion of positive dependency that becomes apparent through rank ordering, and can thus be leveraged by \m{max-rank}. If we compare the copula $C_{\mathbf{X}}$ under \emph{PLOD} to the independent copula $C^{\perp}_{\mathbf{X}}$ evaluated at $\mathbf{q_*}$, it directly follows from \autoref{def:plod} that 
\begin{equation}
    C_{\mathbf{X}}(q_*, \dots, q_*) \geq C^{\perp}_{\mathbf{X}}(q_*, \dots, q_*) \geq 1-\alpha,
\label{eq:copula-plod}
\end{equation}
\ie, $C_{\mathbf{X}}$ stochastically dominates $C^{\perp}_{\mathbf{X}}$. Thus a solution $q_{\max}$ to \autoref{eq:copula-opt-problem} under \emph{PLOD} will be at most as large as a solution to the constraint under independence (\autoref{eq:cdf-indep}), \ie, we have $q_{\max} \leq q^{\perp}_{\max}$ using \m{max-rank}. In contrast the Bonferroni correction only becomes increasingly conservative under dependency (see, \eg, \autoref{fig:global-level-test-samp}), thus no advantage is gained over the solution under independence $q_{\text{Bonf}}$. Together with $q^{\perp}_{\max} \leq q_{\text{Bonf}}$ this implies that \m{max-rank} always performs on par or better than the Bonferroni correction, both under settings of independence and positive dependency. 

We summarize our arguments by restating our proposition from \autoref{sec:theory}, which follows from the above:
\begin{proposition2}
    The solution $q_{\max}$ provided by the \m{max-rank} procedure (\autoref{algo:max-rank}) to \autoref{eq:copula-opt-problem-0} ensures that we have $q_{\max} \leq q^{\perp}_{\max} \leq q_{\text{Bonf}}$ in the case of both independent and positively dependent test statistics.
% \label{prop:better-than-bonf}
\end{proposition2}

\subsection{Using \m{max-rank} for conformal prediction}
\label{subsec:conf-max-rank}

Two key conditions are imposed on any multiple testing correction that integrates with conformal prediction: \emph{(i)} the exchangeability condition on the underlying data is preserved, and \emph{(ii)} \autoref{eq:mht-problem} is corrected in such a way that the conformal coverage guarantee is upheld. We next show that \m{max-rank} satisfies both conditions by looking at its two key operations, namely \emph{(i)} operating in the rank space of nonconformity scores, and \emph{(ii)} applying a $\max$-operation to these ranks. We first address the exchangeability requirement, followed by validity.

\textbf{Preserving exchangeability.} Our arguments rely on work by \cite{ak.kuchibhotla2021}, which establishes connections between exchangeability and rank-based hypothesis testing. Recall our rank definition in \autoref{eq:rank-definition}. For operating in the rank space of nonconformity scores, we make use of the following theorem:
\begin{theorem}[Distribution of ranks \citep{ak.kuchibhotla2021}, Thm. 2]
    For exchangeable random variables $X_1, \dots, X_n$ we have that $$(\mathrm{rank}(X_i; \{X_{1:n}\}): i \in [n]) \sim \mathrm{Unif}(\{\pi: [n] \rightarrow [n]\}),$$ where $\mathrm{Unif}$ is the uniform distribution over all permutations of $[n]$, \ie, each permutation occurs with equal probability $1/n!$.
\label{thm:dist-ranks}
\end{theorem}
It follows that the set of ranks for any underlying set of exchangeable nonconformity scores $S = \{s_1, \dots, s_n\}$ is exchangeable itself, since the ranks are uniformly distributed and do not depend on the distribution of scores. We can leverage \autoref{thm:exch} below to show that particular choices of score functions ensure exchangeable score sets, but here refer to the broader fact that the score set is generally known to be exchangeable since most score functions (\eg, absolute residual scores of the form $|Y_i - \hat{f}(X_i)|$) are symmetric and applied point-wise per (calibration) sample \citep{angelopoulos2023gentle, rj.tibshirani2019, a.fisch2022}. 

For the exchangeability preservation of the $\max$-operation, we now explicitly make use of the following theorem:
\begin{theorem}[Exchangeability under transformations \citep{ak.kuchibhotla2021, d.commenges2003}]
Given a vector of exchangeable random variables $X = (X_1,\dots,X_m) \in \mathcal{X}^m$ and a fixed transformation $G$, we consider $G$ exchangeability-preserving if for each permutation $\pi_1: [m] \rightarrow [m]$ there exists a permutation $\pi_2: [m] \rightarrow [m]$ s.t. $$\forall x \in \mathcal{X}^m : \pi_1 G(x) = G(\pi_2\,x).$$
\label{thm:exch}
\end{theorem}
\vspace{-7mm}
Following \autoref{algo:max-rank}, we denote by $\mathbf{r}_i = (r_{i,1} \cdots r_{i,m}) \in \mathbb{N}^m$ the $i$-th row of the matrix of rank statistics $\mathbf{R}$. The ranks are exchangeable by \autoref{thm:dist-ranks}, and they are random variables since any data randomness propagates through the fixed score function onto the score ranks. If we fix the transformation $G$ to be the $l^{\infty}$-norm, then it follows for its application to any row $\mathbf{r}_i,\,i \in [n]$ that
\begin{equation}
    \pi_1 G(\mathbf{r}_i) = \pi_1 \lVert \mathbf{r}_i \rVert_{\infty} = \pi_1 \underset{1 \le k \le m}{\max} r_{i,k} = \underset{1 \le k \le m}{\max} \pi_2 r_{i, k} = \lVert \pi_2 \mathbf{r}_i \rVert_{\infty} = G(\pi_2 \,\mathbf{r}_i),
\label{eq:exch-ranks}
\end{equation}
satisfying \autoref{thm:exch} and the exchangeability of the $\max$-operation. Intuitively, the $l^{\infty}$-norm is symmetric and permutation-invariant, that is, a permutation $\pi_2$ to its arguments does not change the output permuted by $\pi_1$.

\textbf{Ensuring validity.} We next address the validity requirement, similarly tackling the two key operations of \m{max-rank}. To show that validity and thus \autoref{eq:conf-guarant} is maintained also when operating in the rank space, we employ the following corollary:
\begin{corollary}[\cite{ak.kuchibhotla2021}, Cor. 1]
    Under assumptions of \autoref{thm:dist-ranks} we have that $$\mathbb{P}(\text{rank}(X_n; \{X_{1:n}\}) \le t) = \frac{\lfloor t \rfloor}{n}$$ for $t \in \mathbb{N}$. In addition, the random variable $\text{rank}(X_n; \{X_{1:n}\})/n$ is a valid p-value.
\label{cor:prob-rank}
\end{corollary}
Consider a one-dimensional test sample $(Y_{n+1}, X_{n+1})$, or equivalently, fix a specific $k$-th test dimension. Then we denote the extended set of nonconformity scores by $S' = S \cup \{s_{n+1}\}$, where $s_{n+1}$ is the test sample's score. Using \autoref{eq:rank-definition} and a sorted score set $S'$ observe that $\text{rank}(s_i; S') = i$, \ie, the rank of the $i$-th score is its position index $i$. By \autoref{eq:equiv-of-events} we have that $Y_{n+1} \in \hat{C}(X_{n+1})$ is equivalent to $s_{n+1} \leq \hat{Q}(1-\alpha;S)$, which in the rank space and using the conformal quantile's definition translates to $\text{rank}(s_{n+1}; S') \leq t$, with $t = \left\lceil (n+1)(1-\alpha) \right\rceil$. Using rank exchangeability and \autoref{cor:prob-rank} it follows that
\begin{equation}
    \begin{split}
        \mathbb{P}(Y_{n+1} \in \hat{C}(X_{n+1})) 
        = \mathbb{P}(\text{rank}(s_{n+1}; S') \le t)
        % = \sum_{i=1}^t \mathbb{P}(\text{rank}(s_{n+1}; S') = i)
        = \frac{\lfloor t \rfloor}{n+1}
        = \frac{\lceil(n+1)(1-\alpha)\rceil}{n+1} \ge \frac{(n+1)(1-\alpha)}{n+1} = 1-\alpha,
    \end{split}
\end{equation}
thus ensuring target coverage $(1-\alpha)$. Alternatively, note that valid coverage also directly follows from \m{max-rank}'s FWER-controlling property (\autoref{prop:opt-max-rank}).

For the $\max$-operation, we make a simple argument on stochastic dominance of the quantiles. By definition of the composite empirical null $\mathbf{r}_{\max}$ in \autoref{algo:max-rank} we have for any $k \in [m]$ that $\hat{Q}(1-\alpha; \mathbf{r}_{\max}) \geq \hat{Q}(1-\alpha; \mathbf{r}_k)$ holds. Clearly, a prediction set $\hat{C}_{k}$ for the $k$-th dimension constructed using $\hat{Q}(1-\alpha; \mathbf{r}_{k})$ is then a subset of the prediction set $\hat{C}_{\max}$ in the $k$-th dimension constructed using $\hat{Q}(1-\alpha; \mathbf{r}_{\max})$, \ie, $\hat{C}_{\max} \supseteq \hat{C}_k$. Since by \autoref{eq:equiv-of-events} and validity of operating in the rank space $\hat{C}_k$ is constructed to ensure target coverage, the same coverage guarantee applies to $\hat{C}_{\max}$.

We summarize our results by restating our proposition from \autoref{subsec:conf-max-rank-0}, which follows from the above:
\begin{proposition3}
    The \m{max-rank} procedure (\autoref{algo:max-rank}) provides a multiple testing correction for conformal prediction that preserves exchangeability and ensures valid coverage (\autoref{eq:conf-guarant}).
% \label{prop:max-rank-for-conf}
\end{proposition3}

\section{ALGORITHMIC DETAILS}
\label{app:algo}

The algorithmic descriptions for \cite{westfall1993resampling}'s \m{max-T} and \m{min-P} multiple testing corrections are given in \autoref{algo:max-t} and \autoref{algo:min-p} respectively. The main paper contains the algorithms for split conformal prediction in the general case (\autoref{algo:split-cp}) and our correction \m{max-rank} (\autoref{algo:max-rank}). A simple implementation of \m{max-rank} takes only a few lines of code, and we provide an example in Listing \ref{lst:maxrank}. In this example we simulate perfect correlation ($\rho = 1.0$) and observe how \m{max-rank} exploits the rank ordering and therefore does not adjust the initial significance level (\ie, $\hat{\alpha} = \alpha = 0.1$) which returns the (identical) optimal quantile in each test dimension ($m=5$). In contrast, Bonferroni would adjust to $\hat{\alpha}_{\text{Bonf}} = \alpha / m = 0.02$ in each dimension. For $\rho \approx 0$ the same quantiles as Bonferroni are recovered, whereas $\rho \in (0,1)$ provides tighter quantiles (with varying degree).

\begin{algorithm}[h]
\caption{Multiple testing correction via \m{max-T} (\cite{westfall1993resampling})}
\label{algo:max-t}
\begin{algorithmic}[1]

    \State \textbf{Input:} Hypothesis tests $k=1,\dots,m$ and associated test statistics $(T_1, \dots, T_m) \in \mathbb{R}^m$, a resampling counter $i \in [n]$, and empty vector of resampled test statistics $\mathbf{t}_{\max} \in \mathbb{R}^{n}$.

    \State \textbf{Output:} Adjusted p-values $(P^{\text{adj}}_1, \dots, P^{\text{adj}}_m) \in [0,1]^m$ with FWER control at level $\alpha$.

    \State \textbf{Procedure:} 

    \State For $i=1,\dots,n$:
    
    Resample data, \eg, via permutation or bootstrapping.

    Recompute test statistics for $i$-th resampling and store maximum statistic: $\mathbf{t}_{\max, i} = \max_{1 \leq k \leq m} T_{i,k}$.

    \State Compute adjusted p-value by comparing observed statistics to the maximum statistics under the null:

    $P^{\text{adj}}_k = \frac{1}{n} \sum_{i=1}^{n} \mathds{1}[T_k \leq \mathbf{t}_{\max, i}]$.

    \State Decision rule: $\forall k \in [m]$, we cannot reject the null if $P^{\text{adj}}_k > \alpha$.
    
    \State \textbf{End Procedure}
\end{algorithmic}
\end{algorithm}

\begin{algorithm}[ht]
\caption{Multiple testing correction via \m{min-P} (\cite{westfall1993resampling})}
\label{algo:min-p}
\begin{algorithmic}[1]

    \State \textbf{Input:} Hypothesis tests $k=1,\dots,m$ and associated test statistics $(T_1, \dots, T_m) \in \mathbb{R}^m$, a resampling counter $i \in [n]$, and empty vector of resampled p-values $\mathbf{p}_{\min} \in [0,1]^{n}$.

    \State \textbf{Output:} Adjusted p-values $(P^{\text{adj}}_1, \dots, P^{\text{adj}}_m) \in [0,1]^m$ with FWER control at level $\alpha$.

    \State \textbf{Procedure:} 

    \State For all test statistics, compute the associated unadjusted p-values $(P_1, \dots, P_m) \in [0,1]^m$.

    \State For $i=1,\dots,n$:
    
    Resample data, \eg, via permutation or bootstrapping.

    Recompute p-values for $i$-th resampling and store the minimum p-value: $\mathbf{p}_{\min, i} = \min_{1 \leq k \leq m} P_{i,k}$.

    \State Compute adjusted p-value by comparing observed p-values to the minimum p-values under the null:

    $P^{\text{adj}}_k = \frac{1}{n} \sum_{i=1}^{n} \mathds{1}[P_k \geq \mathbf{p}_{\min, i}]$.

    \State Decision rule: $\forall k \in [m]$, we cannot reject the null if $P^{\text{adj}}_k > \alpha$.
    
    \State \textbf{End Procedure}
\end{algorithmic}
\end{algorithm}
\newpage

\lstset{
    language=Python,                 % Set language
    basicstyle=\ttfamily\footnotesize, % Set font and size
    keywordstyle=\color{blue!90!black}\bfseries,  % Keywords in bold blue
    commentstyle=\color{gray!80!black}\itshape,  % Comments in italic gray
    stringstyle=\color{red!70!black},  % Strings in red
    numberstyle=\tiny\color{gray},     % Line numbers in gray
    numbers=none,                      % Line numbers on the left
    stepnumber=1,                       % Show every line number
    breaklines=true,                    % Line breaking
    frame=lines,                        % Adds a top and bottom border
    rulecolor=\color{black},            % Frame border color
    backgroundcolor=\color{yellow!5},   % Light background for contrast
    tabsize=4,                          % Tab spacing
    showspaces=false, showstringspaces=false, % Hide spaces
    morekeywords={np, axis},             % Add more custom keywords
}

\begin{lstlisting}[caption={Exemplary Python code for the \m{max-rank} correction.}, label={lst:maxrank}]
import numpy as np

alpha = 0.1; n = 1000; m = 5; corr = 1.0 # params

# simulate correlated score matrix
cov = np.full((m, m), corr)
np.fill_diagonal(cov, 1)
S = np.random.multivariate_normal(np.zeros(m), cov, n)

# max-rank correction
R = np.argsort(np.argsort(S, axis=0), axis=0) # column-wise rank matrix
r_max = np.max(R, axis=1) # row-wise max-rank vec(r)_max
q = np.ceil((1 - alpha) * (n + 1)) / n # conformal target coverage level
rank_q = np.quantile(r_max, q, axis=0, method="higher") # quantile r_max
adj_quantile = np.sort(S, axis=0)[rank_q] # final corrected quantiles

print(f"Adjusted significance level: {(1 - rank_q / n):.2f}") # 0.10
print(f"Adjusted quantiles: {adj_quantile}") # e.g. [1.32, 1.32, 1.32, 1.32, 1.32]
\end{lstlisting}

\section{ADDITIONAL BACKGROUND}
\label{app:background}

We provide additional details on FDR control (\autoref{subsec:app-fdr-control}) and the Bonferroni correction (\autoref{subsec:app-fwer-bonferroni}).

\subsection{FDR control}
\label{subsec:app-fdr-control}

In contrast to the stronger notion of FWER control which bounds the probability of committing at least \emph{one} Type-I error, controlling the false discovery rate (FDR) targets a limit on the number of false discoveries \emph{in expectation} across hypothesis tests. This weaker notion of Type-I error control comes at the benefit of a larger fraction of significant discoveries \citep{shaffer1995multiple}. Denote the number of false positives $m_{\text{FP}}$ as before, and the fraction of correct discoveries (\ie, true positives) as $m_{\text{TP}}$. Then, FDR control at significance level $\alpha$ ensures that
\begin{equation}
    \mathbb{E}_{m}(m_{\text{FP}}/(m_{\text{FP}} + m_{\text{TP}})) \leq \alpha,
\label{eq:fdr}
\end{equation}
where the expectation is taken over the $m$ hypotheses. FDR control is appropriate in settings in which a limited proportion of false positives can be tolerated in exchange for a greater chance at significant results, such as in genomics studies \citep{han2009genepos}. Commonly known methods for FDR control include the Benjamini-Hochberg \citep{benjamini1995controlling}, Benjamini-Yekutieli \citep{y.benjamini2001} and Simes corrections \citep{rj.simes1986}\footnote{Simes was originally designed for FWER control, but then adapted to FDR control.}. In contrast to classical corrections with FWER control which assume independence, FDR-controlling procedures are generally more robust to dependencies between tests. Beyond well-known FWER and FDR control, extensions such as $k$-FWER \citep{romano2007control, janson2016familywise} -- which bounds the probability of committing more than $k$ Type-I errors (recovering FWER for $k=1$) -- and false discovery exceedance \citep{genovese2004stochastic} -- which controls tail probabilities of the false discovery proportion (recovering FDR for the expectation) -- have also been explored.

\subsection{FWER control via the Bonferroni correction}
\label{subsec:app-fwer-bonferroni}

We demonstrate how the Bonferroni correction ensures FWER control using a simple union bound argument. Denote for our $m$ hypotheses the corresponding p-values as $P_1, \dots, P_m$. Observe that in the main paper any permutation-based p-values (\ie, on the basis of \autoref{eq:conformal-p-value}) are described as $\hat{P}_1, \dots, \hat{P}_m$ to highlight their data-driven nature, whereas we do not specify here how the Bonferroni p-values are computed. Let us label the fraction of hypotheses that should be correctly marked as non-significant (\ie, true negatives) as $m_{\text{TN}}$. The null is rejected when the respective p-value reaches below significance level, \ie, when we have $P_k \leq \alpha_{\text{Bonf}}$ for some test $k$. Evaluating the risk of incorrectly rejecting a test that belongs to $m_{\text{TN}}$ (\ie, generating a false positive), we obtain that
\begin{equation}
    \text{FWER} = \mathbb{P}\left( \bigcup_{k=1}^{m_{\text{TN}}}(P_k \leq \alpha_{\text{Bonf}})\right) = \mathbb{P}\left(\bigcup_{k=1}^{m_{\text{TN}}}(P_k \leq \frac{\alpha}{m})\right) \leq \sum_{k=1}^{m_{\text{TN}}}\mathbb{P}(P_k \leq \frac{\alpha}{m}) = \sum_{k=1}^{m_{\text{TN}}}\frac{\alpha}{m} = \frac{m_{\text{TN}}}{m}\alpha \leq \alpha.
\label{eq:bonf-fwer}
\end{equation}
We observe that since it follows that FWER $\leq \alpha$, employing $\alpha_{\text{Bonf}} = \alpha/m$ provides valid FWER control.

\section{ADDITIONAL EXPERIMENT DETAILS}
\label{app:experiments}

We give additional experiment details for our simulation study (\autoref{subsec:app-exp-simulation}), multiple testing baselines (\autoref{subsec:app-exp-baselines}), metrics (\autoref{subsec:app-exp-metrics}), and conformal experiments (\autoref{subsec:app-exp-multireg}, \autoref{subsec:app-exp-objdetect}).

\subsection{Simulation study: FWER control and sample size}
\label{subsec:app-exp-simulation}

\begin{figure}[t]
    \centering
    \includegraphics[width=\linewidth]{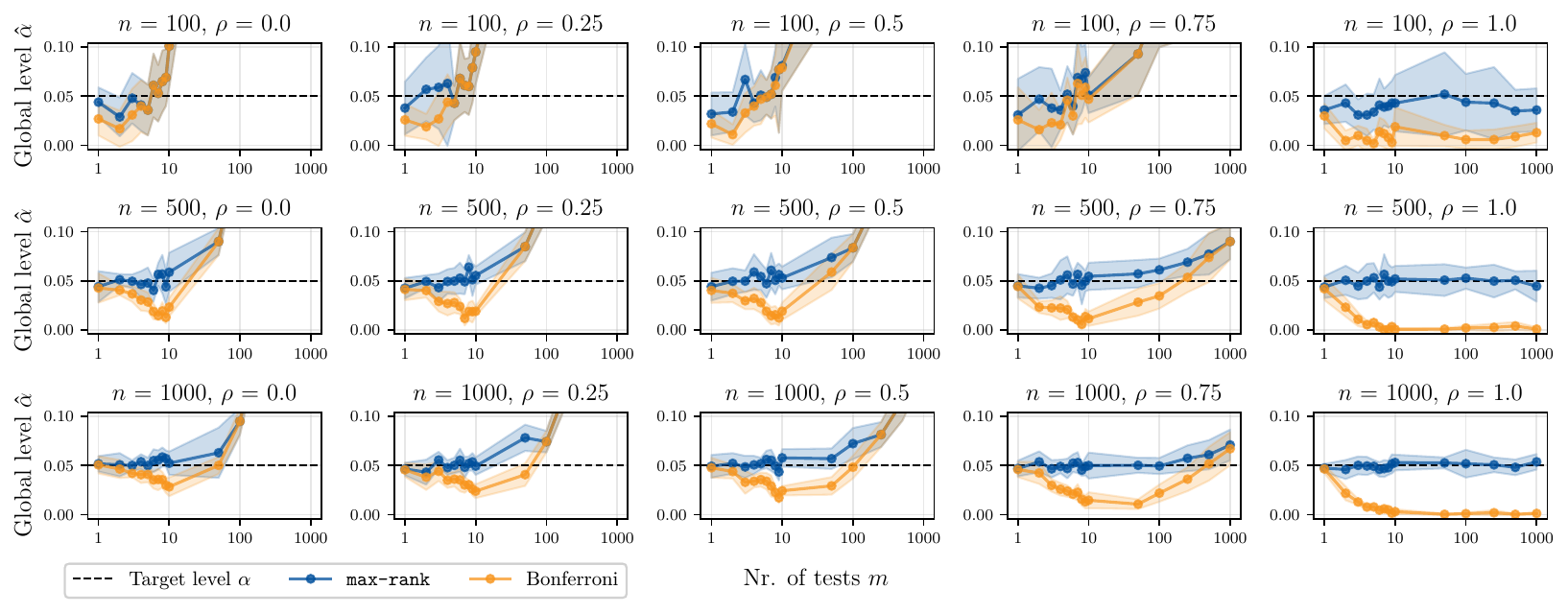}
    \caption{
    We examine \emph{global} adjusted significance levels $\hat{\alpha}$ against the desired FWER control level $\alpha=0.05$ for varying correlations $\rho$, number of tests $m$ and sample size $n$ (shading denotes std. deviation across 10 random trials). Empirical FWER control rates are directly related to the magnitudes of $m$ and $n$, suggesting that larger number of parallel tests require larger sample sizes. We also see that Bonferroni is consistently more conservative than \m{max-rank} across all combinations.
    }
    \label{fig:global-level-test-samp}
\end{figure}

Since \m{max-rank} connects to \cite{westfall1993resampling}'s corrections as a resampling-based method, we anticipate that its data-driven nature requires an adequate ratio of tests counts $m$ to sample sizes $n$ per test (equivalent to the number of resamples or permutations) in order to empirically provide FWER control. Thus, we verify in \autoref{fig:global-level-test-samp} adjusted significance levels for \m{max-rank} and Bonferroni across varying combinations of correlation $\rho$, test counts $m$, and samples $n$. Specifically, we consider $\rho \in \{0, 0.25, 0.5, 0.75, 1.0\}$, $m \in [1, 1000]$, and $n \in \{100, 500, 1000\}$. In particular, this also includes high-dimensional settings where $m > n$.

Our results confirm the anticipated dependencies. Specifically, empirical FWER control is not guaranteed when the ratio of samples to test dimensions is too low, including the settings where $m > n$. These violations affect both \m{max-rank} and Bonferroni, albeit our correction tends to violate sooner since it operates closer to target level. However, FWER control rates improve substantially as both sample size and correlation increases. Intuitively, higher correlation effectively reduces the amount of independent rank order information across tests that needs to be accounted for. Overall, we observe that empirical FWER control is approximately maintained when $n \geq 100 \cdot m$ in the least favorable setting ($\rho = 0$). However, there are cases where this rate can be exceeded (\eg, $n=100$ and $m=5$ holds) or fall short (\eg, $n=1000$ and $m=10$ undercorrects slightly). Finally, throughout all considered combinations Bonferroni is always more conservative than \m{max-rank} (in line with \autoref{prop:better-than-bonf}), and its disadvantage to \m{max-rank} increases substantially as the correlation goes up.

The observed sample requirements seem reasonable given that a reliable empirical null distribution must be established for each test dimension, and are somewhat in the range of calibration set sizes commonly used in conformal prediction. For instance, \cite{angelopoulos2023gentle} recommend at least 1000 calibration samples for a robust conformal procedure even in a single dimension, while \cite{messoudi2021copula} (see Table 2 in their paper) consider varying sample requirements for multi-target regression such as $(n, m) \in \{(2100, 3), (24160, 4), (910, 8)\}$. The impact of overly small sample sizes on results is evident both in their work (\eg, Fig. 3 in their paper) and ours (\autoref{tab:conf-reg}, for the scpf dataset). 

In general, the observed interactions between quantities seem consistent and suggest a meaningful relationship that warrants future theoretical investigation, as mentioned in \autoref{sec:conclusion}. It would be desirable to quantify the expected FWER control achievable for a given significance level $\alpha$, sample size $n$ and number of tests $m$ in analytical form. 

\subsection{Multiple testing correction baselines}
\label{subsec:app-exp-baselines}

We consider the following multiple testing correction baselines in our experiments (\autoref{subsec:conf-exp-reg}): Bonferroni, Bonferroni-Šidák, Gumbel Copula, Empirical Copula, and \m{max-T}. We refer to \autoref{subsec:app-fwer-bonferroni} for the Bonferroni correction and \autoref{methods:relation-to-westfall-young} as well as \autoref{algo:max-t} for the \m{max-T} correction. We comment on the remaining corrections.

\textbf{Bonferroni-\v{S}id\'{a}k.} The correction is a slight adjustment to the classical Bonferroni correction that occasionally improves its performance, but equally imposes an independence assumption for FWER control. An adjusted significance level for target level $\alpha$ is given by $\alpha_{\text{\v{S}id\'{a}k}} = 1 - (1 - \alpha) ^{1/m}$.

\textbf{Gumbel Copula.} The Gumbel copula is a parametric copula which, for some multivariate input $\mathbf{u} = (u_1, \dots, u_m) \in [0,1]^m$, evaluates to
\begin{equation}
    \label{eq:gumbel}
    C_{\text{Gumbel}}(\mathbf{u}) = \exp\left\{ - \sum_{k=1}^{m} (- \ln u_k)^{\theta} \right\}^{1/\theta}.
\end{equation}
It has a light-weight parametric form that only depends on a single parameter $\theta$, which is estimated numerically. \cite{messoudi2021copula} show that under the restriction $u = u_1 = \dots = u_m$ (\ie, $\alpha = \alpha_1 = \dots = \alpha_m$ and thus $\mathbf{u} = (1-\alpha, \dots, 1-\alpha)$ in our setting) a solution to the coverage condition in \autoref{eq:copula-opt-problem-0} is given analytically as $\alpha_{\text{Gumbel}} = 1 - (1 - \alpha)^{1/\sqrt[\theta]{m}}$. Our adjusted significance level is then obtained by plugging in the estimate $\hat{\theta}$ into the equation. We implement both the Gumbel and empirical copulas using the library \m{copulae}\footnote{See \url{https://github.com/danielbok/copulae}}.

\textbf{Empirical Copula.} The empirical copula is the copula-equivalent of an empirical, non-parametric estimator of the joint cumulative distribution. It evaluates to 
\begin{equation}
    \label{eq:emp-copula}
    C_{\text{Emp}}(\mathbf{u}) = \frac{1}{n}\sum_{i=1}^{n}\prod_{k=1}^{m}\mathds{1}[u_{i,k} \leq u_k],
\end{equation}
where $\mathds{1}[\cdot]$ is the indicator function, $u_k$ is a component of $\mathbf{u}$ as above (equivalent to $1-\alpha$), and $u_{i,k}$ is the cumulative probability of the $i$-th sample under its marginal c.d.f. in the $k$-th dimension (estimated empirically), \ie, $u_{i,k} = \hat{F}_{X_k}(x_{i,k})$. We can think of the copula as empirically estimating the fraction of samples jointly satisfying the condition $\mathds{1}[u_{i,k} \leq u_k]$ in all dimensions, since only in such cases does the product evaluate to one. As there is no analytical solution \citep{messoudi2021copula} we have to search through a grid of possible adjusted (global) significance levels $\hat{\alpha} \in (0,1)$, evaluate them under the copula, and select the largest one satisfying target coverage $C_{\text{Emp}}(1-\hat{\alpha}, \dots, 1-\hat{\alpha}) \geq (1-\alpha)$, thus directly addressing the problem in \autoref{eq:copula-opt-problem-0} (note that $\min q_* \Leftrightarrow \max \hat{\alpha}$). 

\subsection{Conformal prediction metrics: empirical coverage and prediction set size}
\label{subsec:app-exp-metrics}

Our approaches are evaluated across key metrics employed in conformal prediction which jointly assess its main desiderata \citep{v.vovk2005, angelopoulos2023gentle}: empirical coverage to assess \emph{validity}, \ie, the coverage target in \autoref{eq:conf-guarant}, and prediction set size to assess \emph{efficiency}. The efficiency of the solution captures its usefulness in practice, since a trivial solution such as an infinite set would meet coverage goals but be of little use. Another popular assessment criterion is \emph{adaptivity}, \ie, the ability to cover equally across different subgroups of the data. Since we are less preoccupied with the nature of our conformal guarantee in this work (marginal \emph{vs.} conditional \emph{vs.} balanced), we do not consider it here.

Moving beyond a single test sample $(X_{n+1},Y_{n+1})$ as frequently denoted in the main paper, let us specify a test set of arbitrary size $n_t$ as $\mathcal{D}_{test} = \{(X_j,Y_j)\}_{j=n+1}^{n+n_t}$. We only consider a test sample covered when all its $k \in [m]$ test dimensions are jointly covered. Thus the average empirical coverage is given by 
\begin{equation}
    \label{eq:metric-cov}
    \text{Coverage} = \frac{1}{n_t\,m} \sum_{j=n+1}^{n+n_t}\mathds{1}\left[\bigcap_{k=1}^{m}(Y_{j,k} \in \hat{C}(X_{j,k}))\right],
\end{equation}
where $\mathds{1}[\cdot]$ is the indicator function. We note that the empirical coverage is a \emph{random} quantity parametrized by a coverage distribution, and can slightly deviate from target coverage $(1-\alpha)$ based on factors such as the size of the calibration data $|\mathcal{D}_{cal}|$ \citep{v.vovk2005}. For efficiency, we compute the average prediction set size across samples and test dimensions as
\begin{equation}
    \label{eq:metric-size}
    \text{Interval Width} = \frac{1}{n_{t}\,m} \sum_{j=n+1}^{n+n_t} \sum_{k=1}^{m}|\hat{C}(X_{j,k})|.
\end{equation}
Note that since our regression experiments in \autoref{subsec:conf-exp-reg} produce prediction sets in the form of intervals, we directly define the metric as prediction \emph{interval width} rather than the more generic term on prediction \emph{set size}. 

\subsection{Conformal multi-target regression}
\label{subsec:app-exp-multireg}

We provide a few more details on our experimental design, which follows closely \cite{messoudi2021copula}. We select a subset of three interesting datasets with varying numbers of regression targets and sample size lengths\footnote{Obtainable at \url{https://github.com/tsoumakas/mulan}}: scpf (website traffic data, $m=3$, 1137 samples), rf1 (river flow data, 9125 samples, $m=8$), and scm1d (product pricing data, 9803 samples, $m=16$). We use the data pre-processing scheme of \cite{messoudi2021copula}\footnote{See \url{https://github.com/M-Soundouss/CopulaConformalMTR}} and similarly train a multi-target random forest regressor using \m{sci-kit learn} \citep{scikit-learn} with minimal hyperparameter tuning. Each dataset is randomly split into 60\% training, 20\% calibration, and 20\% test data, and the model training and conformal procedure is repeated and results averaged across 100 random trials for robust evaluation.

\subsection{Conformal object detection}
\label{subsec:app-exp-objdetect}

Object detection tasks consist of a dual problem wherein both an object's location, as parametrized via it's four bounding box coordinates (thus $m=4$), and the object's class label need to be predicted. We focus on applying conformal prediction to the bounding box prediction task, which forms a regression problem in the (approximately) continuous pixel space of input images. Our experimental design strongly relates to \cite{mukama2024copula} but builds upon our own developed conformal methodology for object detection \citep{timans2024conformalod}. Instead of considering only objects of a single class `car', we instead apply conformal prediction \emph{separately} to a variety of classes. This is easily achieved by filtering the labelled calibration data by class label, and computing class-specific conformal quantiles via \autoref{algo:split-cp} used to build prediction sets for each test sample based on its (unconformalized) class label. Thus, our results in \autoref{fig:bdd100k-res} averaged across multiple classes (and trials) provide a more robust performance evaluation of each multiple testing correction. 

\begin{figure}[!ht]
    \centering
    \includegraphics[width=0.6\linewidth]{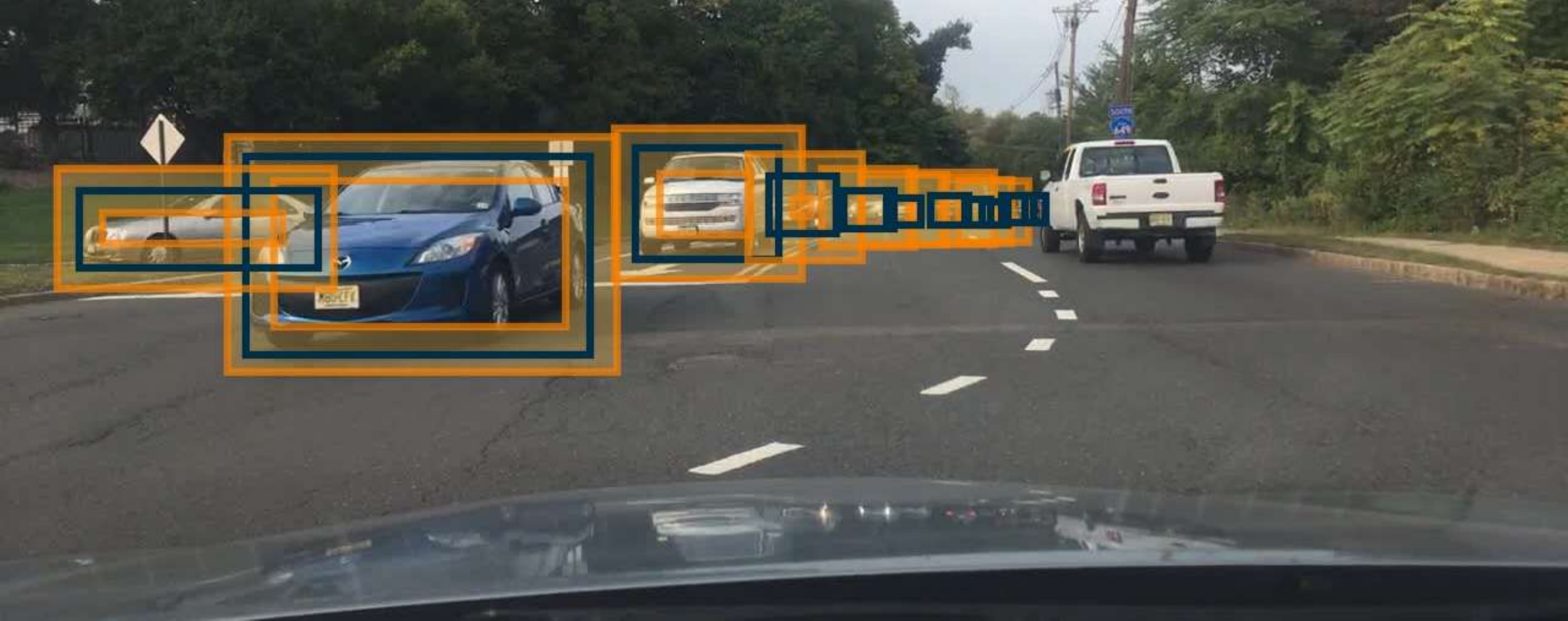}
    \caption{A visual example of the constructed conformal prediction intervals with \m{max-rank} on a test image for the class `car' (and not `truck'). True bounding boxes are in blue, two-sided prediction interval regions are shaded in orange.}
    \label{fig:app-od-visual}
\end{figure}

In particular, we employ the BDD100k dataset \citep{yu2020bdd} in our experiment, and thus consider the classes $\{$person, bicycle, car, motorcycle, bus, truck, traffic light, stop sign$\}$ with varying object counts\footnote{Note that we map classes ‘person’ and ‘rider’ to a single class ‘person’ for simplification.}. We employ a pre-trained object detection model using \m{detectron2} \citep{wu2019detectron2} based on a Faster R-CNN architecture (model type \m{X101-FPN}\footnote{See \url{https://github.com/facebookresearch/detectron2/blob/main/MODEL_ZOO.md}}), and thus do not require training on BDD100k. Instead we split all available labelled data into calibration (70\%) and test sets (30\%), and the conformal procedure is repeated and results are collected across 100 random trials. We additionally report per-class results in \autoref{tab:app-conf-od-perclass} for completeness, affirming that drawn conclusions on \m{max-rank}'s strong performance hold also individually by class. We also display in \autoref{fig:app-od-visual} a visual example of how the constructed prediction intervals around objects look like. We observe that interval regions (shaded in orange) seem reasonably tight and tend to center around the ground truth bounding boxes (in blue). The \m{max-rank} correction ensures that the constructed intervals provide a valid coverage guarantee over the true object locations according to \autoref{eq:conf-guarant}.

\begin{table*}[ht]
    \centering
    \small
    \setlength{\tabcolsep}{6pt}
    \begin{tabular}{llrrr}
         \toprule
         \textbf{Correction} & \textbf{Object class} & Interval Width & Coverage \\
         \\
         \toprule
         Bonferroni & Class avg. &  42.9513 & 0.9134 \\
         & Person & 27.0747 & 0.9131 \\
         & Bicycle & 55.6782 & 0.9160 \\
         & Car & 37.9270 & 0.9135 \\
         & Motorcycle & 60.1595 & 0.9128 \\
         & Bus & 60.4550 & 0.9130 \\
         & Truck & 67.4933 & 0.9122 \\
         & Traffic Light & 15.8933 & 0.9083 \\
         & Stop Sign & 18.9295 & 0.9187 \\
         \midrule
         Bonf.-\v{S}id\'{a}k & Class avg. & 42.1049 & 0.9102 \\
         & Person & 26.5748 & 0.9098 \\
         & Bicycle & 54.7535 & 0.9132 \\
         & Car & 37.0510 & 0.9103 \\
         & Motorcycle & 59.4491 & 0.9103 \\
         & Bus & 58.9098 & 0.9094 \\
         & Truck & 65.9317 & 0.9089 \\
         & Traffic Light & 15.6576 & 0.9048 \\
         & Stop Sign & 18.5114 & 0.9153 \\
         \midrule
         Gumbel Copula & Class avg. & 40.444 & 0.9039 \\
         & Person & 25.8873 & 0.9048 \\
         & Bicycle & 53.1460 & 0.9085 \\
         & Car & 35.4183 & 0.9038 \\
         & Motorcycle & \underline{57.3168} & 0.9015 \\
         & Bus & 56.4559 & 0.9032 \\
         & Truck & 62.3353 & 0.9014 \\
         & Traffic Light & 15.3669 & 0.9006 \\
         & Stop Sign & 17.6255 & 0.9070 \\
         \midrule
         Emp. Copula & Class avg. & \underline{39.9836} & 0.9015 \\
         & Person & \underline{25.2839} & 0.8998 \\
         & Bicycle & \underline{51.5578} & 0.9028 \\
         & Car & \underline{34.5660} & 0.9001 \\
         & Motorcycle & 58.2203 & 0.9053 \\
         & Bus & \underline{55.6454} & 0.9009 \\
         & Truck & 62.1119 & 0.9009 \\
         & Traffic Light & 15.3653 & 0.9006 \\
         & Stop Sign & \underline{17.1182} & 0.9013 \\
         \midrule
         \m{max-T} & Class avg. & 41.1136 & 0.9002 \\
         & Person & 25.2977 & 0.8996 \\
         & Bicycle & 54.9369 & 0.9007 \\
         & Car & 34.6362 & 0.8999 \\
         & Motorcycle & 61.7477 & 0.8988 \\
         & Bus & 57.5168 & 0.9005 \\
         & Truck & \underline{62.0357} & 0.9007 \\
         & Traffic Light & \textbf{15.3196} & 0.9005 \\
         & Stop Sign & 17.4184 & 0.9008 \\
         \midrule
         \m{max-rank} (Ours) & Class avg. & \textbf{39.6185} & 0.8999 \\
         & Person & \textbf{25.2571} & 0.8996 \\
         & Bicycle & \textbf{51.0433} & 0.9005 \\
         & Car & \textbf{34.5276} & 0.8999 \\
         & Motorcycle & \textbf{56.5972} & 0.8988 \\
         & Bus & \textbf{55.3103} & 0.9000 \\
         & Truck & \textbf{61.8884} & 0.9005 \\
         & Traffic Light & \underline{15.3437} & 0.9003 \\
         & Stop Sign & \textbf{16.9807} & 0.8994 \\
         \bottomrule
    \end{tabular}
    \caption{
    Empirical coverage and mean prediction interval width for different corrections ($\alpha=0.1$) on BDD100k for individual classes and the class average. Results are averaged across 100 random trials. \m{max-rank} consistently provides strong results also on a per-class basis (\textbf{best}, \underline{second best} results). We have varying object counts per class, equivalent to the number of per-class calibration samples $n$. \emph{From top to bottom} by class: 44'960 (Class avg.), 35'748, 2032, 280'277, 883, 5151, 10'975, and 21'950.
    } 
    \label{tab:app-conf-od-perclass}
\end{table*}

\end{document}